\documentclass[11pt]{article}\def\today{December 15, 2004}
\usepackage{amssymb}
\usepackage{amsmath}
\usepackage{theorem}
\usepackage{latexsym}
\usepackage{graphicx}
\reversemarginpar
\theoremheaderfont{\bf} 
\theorembodyfont{\sl}
\newtheorem{theo}{Theorem}[section]
{\theorembodyfont{\rm} \newtheorem{defi}[theo]{Definition}}
{\theorembodyfont{\rm} \newtheorem{exa}[theo]{Example}}
{\theorembodyfont{\rm} \newtheorem{rem}[theo]{Remark}}
\newtheorem{prop}[theo]{Proposition}
\newtheorem{cor}[theo]{Corollary}
{\theorembodyfont{\rm} \newtheorem{ass}[theo]{General Assumption}}
\newtheorem{lemma}[theo]{Lemma}
{\theorembodyfont{\rm}}
{\theorembodyfont{\rm}}
\newenvironment{proof}{{\sc Proof:}}{\mbox{}\hfill$\Box$\par}
\newcommand{\eqnref}[1]{~\mbox{$(${\rm \ref{#1}}$)$}}

\renewcommand{\theequation}{\thesection.\arabic{equation}}

\newcommand{\junk}[1]{}
\newcommand{\DS}{\displaystyle}

\newcommand{\N}{{\mathbb N}}
\newcommand{\Q}{{\mathbb Q}}
\newcommand{\F}{{\mathbb F}}

\newcommand{\Z}{{\mathbb Z}}
\newcommand{\cC}{{\mathcal C}}
\newcommand{\cB}{{\mathcal B}}
\newcommand{\cG}{{\mathcal G}}

\newcommand{\cCpol}{\mbox{${\mathcal C}_{\text{pol}}$}}

\newcommand{\rank}{\mbox{\rm rank}\,}

\newcommand{\edge}[2]{\mbox{$-\!\!\!-\!\!\!\longrightarrow$}\hspace{-1.8em}%
\raisebox{1.6ex}{${\scriptscriptstyle (\!\frac{\,{#1}\,}{#2}\!)}$}\hspace{.8em}}

\newcommand{\im}{\mbox{\rm im}\,}

\newcommand{\dist}{\mbox{\rm dist}}
\newcommand{\wt}{\mbox{\rm wt}}

\newcommand{\T}{\mbox{$\!^{\sf T}$}}

\newcommand{\Flaurent}{\mbox{$\F(\!(z)\!)$}}
\newcommand{\Fpower}{\mbox{$\F[\![z]\!]$}}

\renewcommand{\v}{\mbox{$\mathfrak{v}$}}

\newcommand{\Smalltwomat}[2]{\mbox{\footnotesize{$\begin{pmatrix}{#1}\\{#2}\end{pmatrix}$}}}

\newcounter{abc}
\newcounter{def}
\newenvironment{romanlist}{\begin{list}{(\roman{abc})\hfill}{\usecounter{abc}
     \topsep-1.4ex \labelwidth.7cm \leftmargin.7cm \labelsep0cm
     \rightmargin0cm \parsep0ex \itemsep.6ex
     \partopsep1.6ex}}{\end{list}}
\newenvironment{alphalist}{\begin{list}{(\alph{abc})\hfill}{\usecounter{abc}
     \topsep-1.4ex \labelwidth.7cm \leftmargin.7cm \labelsep0cm
     \rightmargin0cm \parsep0ex \itemsep.6ex
     \partopsep1.6ex}}{\end{list}}
\newenvironment{arabiclist}{\begin{list}{(\arabic{abc})\hfill}{\usecounter{abc}
     \topsep-1.4ex \labelwidth.7cm \leftmargin.7cm \labelsep0cm
     \rightmargin0cm \parsep0ex \itemsep.6ex
     \partopsep1.6ex}}{\end{list}}

\title{On the Weight Distribution of Convolutional Codes}
\date\today
\author{Heide Gluesing-Luerssen\footnote{
       University of Groningen, Department of Mathematics, P.~O.~Box 800,
       9700 AV Groningen, The Netherlands; gluesing@math.rug.nl}
       }
\begin{document}
\maketitle

\begin{abstract}
\noindent
Detailed information about the weight distribution of a convolutional code is
given by the adjacency matrix of the state diagram associated with a controller 
canonical form of the code.
We will show that this matrix is an invariant of the code.
Moreover, it will be proven that codes with the same adjacency matrix have the 
same dimension and the same Forney indices and finally that for one-dimensional 
binary convolutional codes the adjacency matrix determines the code uniquely up 
to monomial equivalence.
\end{abstract}



\section{Introduction}
The weight distribution of a code forms an important parameter containing a lot of
information about the code for practical as well as theoretical purposes. 
Quite some research in block code theory has been devoted to the investigation of
the weight distribution and to weight preserving maps.
The most famous results in this area are certainly the Duality Theorem of MacWilliams 
which presents a transformation between the weight distributions of a given code and 
its dual as well as MacWilliams' Equivalence Theorem stating that two isometric codes 
are monomially equivalent. 

In this paper we will address these topics for convolutional codes. 
In~\cite{SM77} it has been shown that there cannot exist a duality theorem for 
the weight distribution of convolutional codes and their duals, if the weight 
distribution is defined to be enumerating the atomic codewords. 
The authors present two binary one-dimensional codes having the same weight distribution 
while their duals have different ones. 
We will present this example at the end of Section~\ref{S-compl.inv}.
It indicates that the thus defined weight distribution contains too little 
information about the code when it comes to duality.
In this paper we will concentrate on a different type of weight distribution 
containing considerably more information.
This is the adjacency matrix of the state diagram associated with the controller 
canonical form of a minimal encoder.
Since for block codes this matrix reduces to the usual weight distribution, it can 
be regarded as a generalization of the latter to convolutional codes.
Then the following issues arise. 
First of all, it needs to be clarified whether this matrix is an 
invariant of the code, i.~e., does not depend on the choice of the minimal encoder.
Since at any rate the adjacency matrix depends on the chosen ordering on the state 
space, it is reasonable to factor out the effect of the ordering. 
This brings us to the notion of generalized adjacency matrix for which then we can
answer the question above in the affirmative in Section~\ref{S-AdjUnique}. 
Secondly and more interestingly, we will approach the problem of how much information 
about the code does the generalized adjacency matrix contain. 
More precisely, what is the relation between two codes sharing the same generalized 
adjacency matrix?
In Section~\ref{S-compl.inv} we will first show that two such codes always have the
same Forney indices. 
Thereafter, we will restrict ourselves to the most important class of convolutional 
codes from a practical point of view, the one-dimensional binary codes. 
With the aid of MacWilliams' Equivalence Theorem for block codes we will show that such 
codes share the same generalized adjacency matrix if and only if they are monomially 
equivalent. 
One should bear in mind that this result cannot be expected for 
general codes, since it is not even true for higher-dimensional binary block codes. 
However, as a consequence we obtain that if two binary one-dimensional convolutional 
codes share the same generalized adjacency matrix then so do their dual codes. 
This shows in particular that the example in~\cite{SM77} mentioned above does not 
apply if we consider the generalized adjacency matrix rather than the weight 
distribution of the atomic codewords.
The result just mentioned tempts us to conjecture that two codes have the same
generalized adjacency matrix if and only if this is true for the dual codes, or, more 
explicitly, that there might be a MacWilliams duality for the generalized adjacency 
matrices of convolutional codes.
Indeed, in~\cite{Ab92} in a totally different form a MacWilliams tranformation rule has 
been established for convolutional codes of the smallest kind, that is, with overall 
constraint length being one.
We will present this transformation in\eqnref{e-MacWtrafo} at the end of this paper.
Proving or disproving the conjecture for general codes, however, appears to be quite a 
difficult problem and has to remain open for future research.

In the next two sections we will introduce the material as necessary for deriving our 
results.
We will discuss the controller canonical form of an encoder
matrix and the associated state diagram along with its adjacency matrix. 
This will in particular bring us to a state-space description of convolutional codes as 
introduced first by Massey and Sain~\cite{MaSa67} and discussed in many other papers, see,
e.~g., \cite{Fo73,McE98}.
Since the results as we will need them later on are somewhat spread over the literature and 
proofs are not always easily accessible, we think it is  
worthwhile presenting the material of the next two sections as self-contained as possible, 
even though most of the results are not new.
Furthermore, as opposed to the existing literature we will give a purely matrix theoretic approach. 
Following McEliece~\cite{McE98a} we will introduce the notions of atomic and molecular 
codewords and show how the corresponding weight distribution can be derived, theoretically, from the 
adjacency matrix of the state diagram.
Other methods of computing the weight distribution are Viterbi's method, 
see~\cite{Vi71} or \cite[Sec.~3.10]{JoZi99}, or Mason's gain formula as 
described in \cite[Sec.~10.2]{LC83}.
For us McEliece's approach has the advantage to be fully in matrix theoretic terms.
This way we will obtain a consistent presentation throughout the whole paper which 
also bears the hope that we can bring this interesting topic closer to the 
mathematical community. 
Moreover, his method leads us directly to the adjacency matrix which will be the 
central object of our studies in the last two sections.
At the end of Section~\ref{S-3}, we will also briefly discuss two types of distance 
parameters which have been introduced in a totally different context for convolutional codes, 
the extended row distances and the active burst distances. 
Both are closely related to the notions of atomic and molecular codewords and, as we will see, 
therefore also appear in the weight distribution.
Sections~\ref{S-AdjUnique} and~\ref{S-compl.inv} contain the new results as described 
in the previous paragraph.
We will close the paper with some open problems in Section~\ref{S-OP}.

We end the introduction with presenting the basic notions of convolutional coding 
theory.
Throughout the paper the symbol~$\F$ stands for any finite field while~$\F_q$ always 
denotes a field with~$q$ elements.
The ring of polynomials and the field of formal Laurent series over~$\F$ are
given by $\F[z]=\big\{\sum_{j=0}^Nf_jz^j\,\big|\,N\in\N_0,\,f_j\in\F\big\}$
and $\Flaurent=\big\{\sum_{j=l}^{\infty}f_jz^j\,\big|\,l\in\Z,\,f_j\in\F\big\}$.
The following definition of a convolutional code is standard.

\begin{defi}\label{D-CC}
Let $\F=\F_q$.
A {\em convolutional code\/} with parameters $(n,k,\delta)_q$ is a $k$-dimensional 
subspace~$\cC$ of the vector space $\Flaurent^n$ of the form 
$\cC=\im G:=\big\{uG\,\big|\, u\in\Flaurent^k\big\}$
where~$G$ is a matrix in $\F[z]^{k\times n}$ that is {\em basic}, i.~e. there exists some matrix 
$\tilde{G}\in\F[z]^{n\times k}$ such that $G\tilde{G}=I_k$, and satisfies 
$\delta=\max\{\deg\gamma\mid \gamma$ is a $k$-minor of $G\}$.
We call $G$ a {\sl generator matrix\/} or {\em encoder\/} and~$\delta$ the 
{\sl overall constraint length\/} of the code~$\cC$.
\end{defi}
Notice that, by definition, a generator matrix is always polynomial and has a
polynomial right inverse.
This implies that in the situation of Definition~\ref{D-CC} the polynomial
codewords belong to polynomial messages, i.~e.
$\cCpol:=\cC\cap\F[z]^n=\big\{uG\,\big|\,u\in\F[z]^k\big\}$.
In other words, the generator matrix is delay-free and non-catastrophic.
As a consequence, a convolutional code is always uniquely determined by its polynomial
part.
Precisely, if $\cC=\im G$ and $\cC'=\im G'$ where
$G,\,G'\in\F[z]^{k\times n}$ are both basic, then
\begin{equation}\label{e-cpolyunique}
  \cC=\cC'\Longleftrightarrow \cC\cap\F[z]^n=\cC'\cap\F[z]^n
  \Longleftrightarrow G'=UG\text{ for some }U\in Gl_k(\F[z]).
\end{equation}
This also shows that the overall constraint length of a code
does not depend on the choice of the generator matrix.

It is well-known~\cite[Thm.~5]{Fo70} or~\cite[p.~495]{Fo75} that each code
has a minimal generator matrix in the sense of the next definition and that such a matrix can 
be obtained constructively from a given generator matrix, see~\cite[Sec.~4]{Fo75}.
For a polynomial vector $v\in\F[z]^n$ we define $\deg v$ to be the maximum degree 
of its entries and, as usual, we put $\deg0=-\infty$.
\begin{defi}\label{D-min}
Let $G\in\F[z]^{k\times n}$ be a basic matrix with overall constraint length~$\delta$
and let $\nu_1,\ldots,\nu_k$ be the degrees of the rows of~$G$.
We say that $G$ is {\sl minimal\/} if $\delta=\sum_{i=1}^k\nu_i$.
In this case, the row degrees of~$G$ are uniquely determined by the code
$\cC:=\im G\subseteq\Flaurent^n$. They are called the {\sl Forney indices\/}
of~$\cC$.
The maximal Forney index is called the {\em memory\/} of the code.
\end{defi}
From the above it follows that a convolutional code with parameters $(n,k,\delta)$ 
has a constant generator matrix if and only if $\delta=0$. 
In that case the code can be regarded as an $(n,k)$ block code.

The definition of weight and distance in convolutional coding theory is straightforward.
For a polynomial vector $v\in\F[z]^n$ with $\deg(v)=N$ and for $j=0,\ldots,N$ we define 
$v_j\in\F^n$ to be the vector coefficient of $z^j$ in~$v$. 
Then the {\em weight\/} of~$v$ is defined to be 
$\wt(v)=\sum_{j=0}^N\wt(v_j)$ where $\wt(v_j)$ denotes 
the Hamming weight of $v_j\in\F^n$.
The {\em (free) distance\/} of a code $\cC\subseteq\Flaurent^n$ is given as
$\dist(\cC):=\min\big\{\wt(v)\,\big|\, v\in\cCpol,\;v\not=0\big\}$.

\section{Controller canonical form and state diagram} \label{S-2}
\setcounter{equation}{0}
In this section we introduce the matrix representation of the controller canonical form 
for a given encoder matrix~$G$.
The interpretation of this form has been discussed in detail in \cite[Sec.~II]{Fo73},
\cite{Fo70}, and \cite[Sec.~2.1]{JoZi99}.
The relation to the encoding process will be made clear below.
The results of this section are in essence well-known from the references above and 
other coding literature. 
However, since we were not able to find detailed references including proofs for all 
results we think it is worthwhile to summarize them with strict matrix theoretical 
proofs.
We will make heavy use of these results in later sections.
\begin{defi}\label{D-ABCD}
Let $G=(g_{ij})\in\F[z]^{k\times n}$ be a generator matrix with row degrees
$\gamma_1,\ldots,\gamma_k$ and let $g_{ij}=\sum_{\nu=0}^{\gamma_i}g_{ij}^{(\nu)}z^{\nu}$.
Put $\gamma=\sum_{i=1}^k\gamma_i$ and asume $\gamma>0$.
For $i=1,\ldots, k$ define
\[
  A_i=\begin{pmatrix} 0\!\!&1\!\!\!& & \\ & &\ddots\!\!& \\& & &1\\ & & &0\end{pmatrix}
      \in\F^{\gamma_i\times\gamma_i},\;
  B_i=\begin{pmatrix}1&0&\cdots&0\end{pmatrix}\in\F^{\gamma_i},\;
  C_i=\begin{pmatrix}g_{i1}^{(1)}&\cdots&g_{in}^{(1)}\\
                     g_{i1}^{(2)}&\cdots&g_{in}^{(2)}\\
                    \vdots       &      &\vdots\\
                     g_{i1}^{(\gamma_i)}&\cdots&g_{in}^{(\gamma_i)}
      \end{pmatrix}\in\F^{\gamma_i\times n}
\]
and put 
\[
   A=\begin{pmatrix} A_1\!\!& &  &\\ &A_2\!\!&   &\\ & &\ddots &\\& & &\!\!A_k\end{pmatrix}
   \in\F^{\gamma\times\gamma},\;
   B=\begin{pmatrix} B_1\!\!& & &\\ &B_2\!\!&   &\\ & &\ddots &
   \\& & &\!\!B_k\end{pmatrix}\in\F^{k\times\gamma},\;
   C=\begin{pmatrix}C_1\\C_2\\ \vdots\\C_k\end{pmatrix}\in\F^{\gamma\times n}
\]
as well as $D=\big(g_{ij}^{(0)}\big)\in\F^{k\times n}$.
In case $\gamma_i=0$ for some~$i$, the corresponding block is missing and in~$B$ a 
zero row occurs.
We call $(A,B,C,D)$ the {\em controller canonical form\/} of the code~$\cC=\im G$.
\end{defi}
Let us first investigate the properties of~$G$ in terms of its controller canonical
form.
\begin{lemma}\label{L-ABCD}
Let $G,\,A,\,B,\,C$, and $D$ be as in Definition~\ref{D-ABCD}. 
Then
\begin{arabiclist}
\item $G=B(z^{-1}I-A)^{-1}C+D=zB(I-zA)^{-1}C+D$.
\item $G$ is minimal in the sense of Definition~\ref{D-min} if and only if 
      $\rank[A,\,C]=\gamma$.
\end{arabiclist}
\end{lemma}
\begin{proof}
(1) It is easy to see that 
\begin{equation}\label{e-B(zI-A)inverse}
  B(z^{-1}I-A)^{-1}=
   \left(\!\!\begin{array}{ccccccccccccc}
      z&z^2&\cdots&z^{\gamma_1}& &   &      &            &      &  &  &       &    \\
       &   &      &            &z&z^2&\cdots&z^{\gamma_2}&      &  &  &       &    \\
       &   &      &            & &   &      &            &\ddots&  &  &       & \\
       &   &      &            & &   &      &            &\qquad&z &z^2&\cdots&z^{\gamma_k}
   \end{array}\!\!\right),
\end{equation}
where a zero row occurs if $\gamma_i=0$.
From this and the definition of all matrices involved we obtain
$B(z^{-1}I-A)^{-1}C=
  \Big(\sum_{\nu=1}^{\gamma_i}g_{ij}^{(\nu)}z^{\nu}\Big)_{i=1,\ldots,k\atop
  j=1,\ldots,n}=G-D$.
This shows the first identity of~(1).
If $\gamma_i=0$, then the $i$-th row of~$B(z^{-1}I-A)^{-1}C$ is zero, and the assertion is correct,
too.
The second equality follows easily from the first one.
\\
(2) Using the fact that~$G$ is minimal if and only if the highest coefficient matrix 
$\big(g_{ij}^{\gamma_i}\big)_{i,j}\in\F^{k\times n}$ has rank~$k$ (see \cite[p.~495]{Fo75}),
this part follows directly from the definition of~$A$ and~$C$.
\end{proof}

It is well-known that the encoding process of the convolutional code generated by~$G$ 
can be described by a linear shift register. 
The matrix version of this is exactly the controller canonical form along
with the corresponding dynamical system as given in part~(1) of the next theorem.  
We will give the interpretation right after the proof. 
\begin{theo}\label{T-ABCD}
Let $G\in\F[z]^{k\times n}$ be a generator matrix and $A,\,B,\,C$, and $D$ be as in 
Definition~\ref{D-ABCD}.
Let $u=\sum_{t\geq0}u_tz^t\in\Flaurent^k$ and $v=\sum_{t\geq0}v_tz^t\in\Flaurent^n$ and
define $x=uB(z^{-1}I-A)^{-1}=zuB(I-zA)^{-1}\in\Flaurent^{\gamma}$.
\begin{arabiclist}
\item Then $x=\sum_{t\geq1}x_tz^t$ (i.~e., $x_0=0$) and
      \[
          v=uG\Longleftrightarrow
          \left\{\begin{array}{rcl} z^{-1}x&=&xA+uB\\v&=&xC+uD\end{array}\right\}
          \Longleftrightarrow
          \left\{\begin{array}{rcl} x_{t+1}&=&x_tA+u_tB\\v_t&=&x_tC+u_tD\end{array}
          \;\text{ for all }t\geq0.\right\}
      \]
\item If $u=(u^{(1)},\ldots,u^{(k)})\in\F[z]^k$, i.~e.~$u$ is polynomial, then $x\in\F[z]^{\gamma}$ and
      \[\deg x=\max\big\{\gamma_i+\deg u^{(i)}\,\big|\,i=1,\ldots,k,\,
	            \gamma_i\not=0\big\}.
       \]
       Moreover, if~$G$ is minimal, then
       \[
          \deg(uG)=\max\big\{\gamma_i+\deg u^{(i)}\,\big|\,i=1,\ldots,k\big\}.
       \]
\end{arabiclist}
We call $x_t\in\F^{\gamma}$  as in~(1) the state of the encoder at time~$t$ given that the input
is~$u$ as above.
The space $\F^{\gamma}$ is called the state space of the encoder~$G$.
\end{theo}
The state-space realization $x_{t+1}=x_tA+u_tB,\,v_t=x_tC+u_tD$ has been introduced 
in~\cite{MaSa67} and has also been discussed in~\cite{Fo73,McE98}.
It is different, however, from the state-space system used in 
\cite{RSY96,RoYo99,RoSm99,Ro01}. 
In those papers the codeword is made up by the combined input and output, while in our case
the codeword coincides with the output of the system.
\\[1ex]
\begin{proof}
(1) From Lemma~\ref{L-ABCD}(1) we have
$v=uG\Longleftrightarrow v=u(B(z^{-1}I-A)^{-1}C+D)$ and using the definition for~$x$
the first equivalence follows.
The other one follows by equating like powers of~$z$.
\\
(2) From\eqnref{e-B(zI-A)inverse} we have
$x=(u^{(1)}z,\ldots,u^{(1)}z^{\gamma_1},\ldots,u^{(k)}z,\ldots,u^{(k)}z^{\gamma_k})$
where for $\gamma_i=0$ the corresponding block is missing.
From this the first assertion follows. The second one is one of the well-known
characterizations of minimal matrices, see \cite[p.~495]{Fo75}.
\end{proof}
Notice that if all row indices $\gamma_i$ are nonzero, then the matrix~$B$ has
full row rank and $\deg x\geq \deg u$. If $\gamma_i=0$ for some~$i$, then $\ker B\not=0$
and the inequality $\deg x<\deg u$ might occur.
In any case one has $\deg x\leq \deg(uG)$.

Obviously the dynamical equations $x_{t+1}=x_tA+u_tB,\,v_t=x_tC+u_tD$ describe the
input-state-output behavior of the canonical shift register realization of the encoder~$G$.
The inputs at time~$t$ are given by the sequence $u_t$, the state vectors $x_t\in\F^{\gamma}$
represent the contents of the memory elements of the register at time~$t$ and~$v_t$
is the output at that time.
Part~(1) above tells us in particular $x_0=0$, which is the usual assumption that the  
shift register is empty at the beginning of the encoding process.
Part~(2) shows that if the input sequence is finite, then the memory is finite, too.
It is zero after a certain number of steps depending on the length of the~$k$ different 
memory series and on when the entering input sequence is zero.

\begin{exa}\label{E-CCF}
Let $q=16$ and 
\[
   G=\begin{bmatrix}
       \alpha+\alpha z+z^2 & \alpha^6+\alpha z+\alpha^{10}z^2
          & \alpha^{11}+\alpha z+\alpha^5 z^2\\
       1+z & \alpha^{10}+\alpha^5 z & \alpha^5+\alpha^{10}z
   \end{bmatrix}\in\F_{16}[z]^{2\times 3}
\]
where $\alpha^4+\alpha+1=0$.
Then the multiplication
\[
  uG=\sum_{t\geq0}u_tz^tG=\sum_{t\geq0}
       \underbrace{\big(u^{(1)}_t,\,u^{(2)}_t\big)}_{u_t}z^t\cdot G
     =\sum_{t\geq0}
       \underbrace{\big(v^{(1)}_t,\,v^{(2)}_t,\,v^{(3)}_t\big)}_{v_t}z^t
     =\sum_{t\geq0}v_tz^t=v
\]
is realized by the following linear shift register, shown at time~$t$.
\\[-4ex]
\mbox{}\hspace*{.2cm}
\includegraphics[height=19cm]{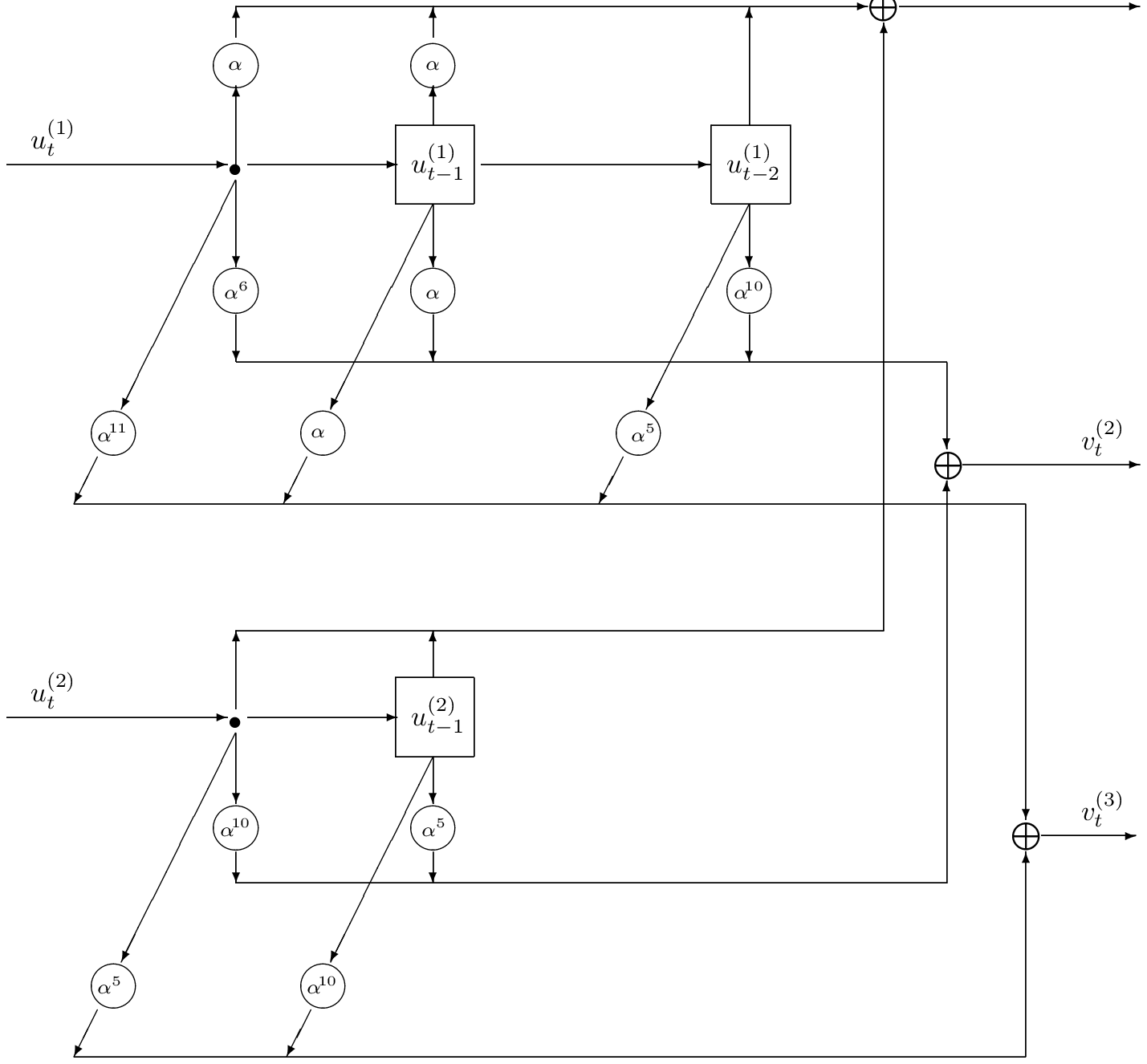}

\vspace*{-9.5cm}
\noindent The controller canonical form is given by
\[
  A=\begin{pmatrix}0&1&0\\0&0&0\\0&0&0\end{pmatrix},\
  B=\begin{pmatrix}1&0&0\\0&0&1\end{pmatrix},\
  C=\begin{pmatrix}\alpha&\alpha&\alpha\\1&\alpha^{10}&\alpha^5\\
                   1&\alpha^5&\alpha^{10}\end{pmatrix},\
  D=\begin{pmatrix}\alpha&\alpha^6&\alpha^{11}\\1&\alpha^{10}&\alpha^5\end{pmatrix}.
\]
The dynamical equations $x_{t+1}=x_tA+u_tB,\,v_t=x_tC+u_tD$ with the state
at time~$t$ being $x_t:=(u^{(1)}_{t-1},u^{(1)}_{t-2},u^{(2)}_{t-1})$ describe exactly the 
input-state-output behavior of the shift register.
\end{exa}

From now on we will always assume that the generator matrix
$G\in\F[z]^{k \times n}$ of the code~$\cC\subseteq\Flaurent^n$ is 
minimal with Forney indices $\gamma_1,\ldots,\gamma_k$ and overall constraint 
length~$\gamma>0$.
Moreover, we define the matrices $A,\,B,\,C$, and~$D$ as in
Definition~\ref{D-ABCD}.

\begin{lemma}\label{L-concatenated}
Let $u\in\F[z]^k$ and $v:=uG\in\cCpol$. Assume $v_0\not=0$ and let $\deg v=N>0$.
Put $x=uB(z^{-1}I-A)^{-1}\in\F[z]^\gamma$.
Choose $L\in\{1,\ldots,N\}$. Then the following are equivalent.
\begin{romanlist}
\item $x_L=0$,
\item $v=\tilde{v}+\hat{v}$ where $\tilde{v},\,\hat{v}\in\cCpol\backslash\{0\}$ and
      $\deg\tilde{v}< L,\,\hat{v}\in z^L\F[z]^n$.
\end{romanlist}
\end{lemma}
This lemma simply states that if the shift register is back to the zero state (at
time $t=L$), then one may regard the information before and after that time instance as two
separate information messages and the associated codewords as two separate
codewords.
\\[1ex]
\begin{proof}
Recall that $\deg x\leq \deg v=N$.
\\
``(i) $\Rightarrow$ (ii)'':
Put $\tilde{x}=\sum_{t=0}^{L-1}x_tz^t$ and $\hat{x}=\sum_{t=L+1}^Nx_tz^t$.
If $L=N$, put $\hat{x}=0$.
Then $x=\tilde{x}+\hat{x}$ and $uB=\tilde{x}(z^{-1}I-A)+\hat{x}(z^{-1}I-A)$.
Writing $u=\tilde{u}+\hat{u}$ where $\deg\tilde{u}<L$ and $\hat{u}\in z^L\F[z]^k$,
we obtain $\deg(\tilde{u}B)<L$ and $\hat{u}B\in z^L\F[z]^\gamma$.
Therefore, $\tilde{u}B=\tilde{x}(z^{-1}I-A)$ and $\hat{u}B=\hat{x}(z^{-1}I-A)$.
Now put $\tilde{v}:=\tilde{u}G=\tilde{u}(B(z^{-1}I-A)^{-1}C+D)=\tilde{x}C+\tilde{u}D$
and $\hat{v}:=\hat{u}G=\hat{x}C+\hat{u}D$.
Then it is easy to see that~(ii) is satisfied.
\\
``(ii) $\Rightarrow$ (i)'':
Let $\tilde{v}=\tilde{u}G$ and $\hat{v}=\hat{u}G$, hence $v=(\tilde{u}+\hat{u})G$.
Then basicness of~$G$ implies $\hat{u}\in z^L\F[z]^k$.
Moreover, since~$G$ is a minimal, we have
\[
  \deg\tilde{v}=\max\{\deg\tilde{u}^{(i)}+\gamma_i\mid i=1,\ldots,k\},
\]
where $\tilde{u}=(\tilde{u}^{(1)},\ldots,\tilde{u}^{(k)})$, see
\cite[p.~495]{Fo75}.
Thus the assumption $\deg\tilde{v}<L$ implies $\deg\tilde{u}^{(i)}<L-\gamma_i$ for all
$i=1,\ldots,k$.
Now, $x=\tilde{u}B(z^{-1}I-A)^{-1}+\hat{u}B(z^{-1}I-A)^{-1}$ and from\eqnref{e-B(zI-A)inverse}
we obtain that $\deg(\tilde{u}B(z^{-1}I-A)^{-1})<L$ and
$\hat{u}B(z^{-1}I-A)^{-1}\in z^{L+1}\F[z]^\gamma$.
Thus $x_L=0$.
\end{proof}

The above gives rise to the distinction of codewords into those which are the sum of
two non overlapping codewords and those which are not.
For counting weights it will be advantageous to make an even finer distinction.
We will introduce this only for polynomial codewords.
The generalization to infinite codewords is obvious.
\begin{defi}\label{D-atomic}
Let $\cC=\im G$ and $v\in\cCpol$ such that $v_0\not=0$. Let $L\in\N$.
\begin{alphalist}
\item The codeword~$v$ is called {\em concatenated at time\/}~$t=L$ if
      \[
          v=\tilde{v}+\hat{v}\text{ where } \tilde{v},\,\hat{v}\in\cCpol\backslash\{0\},\
          \deg\tilde{v}=L-1,\,\hat{v}\in z^L\F[z]^n.
      \]
      If additionally, $v_L\not=0$, we call~$v$ {\em tightly concatenated at time\/}~$t=L$.
\item We call the codeword {\em concatenated\/} if it is concatenated at some time
      instance $t=L$. We call it {\em tightly concatenated\/} if each of its
      concatenations is tight.
\item If~$v$ is not concatenated, then~$v$ is called {\em atomic}.
\item If~$v$ is tightly concatenated or atomic, then~$v$ is also called {\em molecular}.
\end{alphalist}
\end{defi}
Parts of this definition can also be found in~\cite{McE98a}. 
Several comments are in order.
First of all, we consider only polynomial codewords that start at time $t=0$, i.~e.,
$v_0\not=0$.
This is certainly no restriction when it comes to computing the weight.
Secondly, it is obvious that each such codeword is the concatenation of atomic 
codewords. 
Thirdly, from Lemma~\ref{L-concatenated} we know that if~$v$ is concatenated at time 
$t=L$ then the state at time $t=L$ satisfies~$x_L=0$.
In this case, the dynamical equations in Theorem~\ref{T-ABCD}(1) show that~$v$ is tightly concatenated
at time $t=L$ if and only if $u_L\not=0$ (since the matrix~$D$ is right invertible).
Thus for a non-tightly concatenated codeword the shift register is zero and the input is zero
for at least one time instance before nonzero input is entering again.
For a tightly concatenated codeword the shift register is zero, and there is immediately
nonzero input being fed into the system.
If~$v$ is concatenated at time $t=L$, but not tightly concatenated, then we have
$x_L=x_{L+1}=0$.
However, if~$v$ is tightly concatenated at time $t=L$, then it also might happen that 
$x_L=x_{L+1}=0$.
This is because the matrix~$B$ might have a nontrivial kernel.
All this is best visualised by using the state diagram.
It will be advantageous to define it in such a way that it only captures the
molecular codewords.
\begin{defi}\label{D-SD}\
\begin{alphalist}
\item Let $s:=q^{\gamma}$ and write
      $\F^{\gamma}=\{X_0, X_1,\ldots,X_{s-1}\}$ in an
      arbitrary ordering such that $X_0=0$.
      We define $\F^{\gamma}$ to be the {\em state space\/} of the encoder~$G$ and 
      the {\em state diagram\/} of~$G$ as the labeled directed graph 
      given by the vertex set $\{X_0,X_1,\ldots,X_{s-1}\}$ and the set of edges
      \[
         \Big\{X_i\edge{u}{v}X_j\,\Big|\,
          u\in\F^k,\,v\in\F^n:\; X_j=X_iA+uB,\,v=X_iC+uD,\,(X_i,u)\not=(0,0)\Big\}
      \]
\item A {\em path of length~$l$\/} is a sequence of edges of the form
      \begin{equation}\label{e-path}
         X_{i_0}\edge{u_0}{v_0}X_{i_1}\edge{u_1}{v_1}X_{i_2}\edge{u_2}{v_2}\ldots\ldots\
         \mbox{$-\!\!\!-\!\!\!\longrightarrow$}\hspace{-2.5em}
         \raisebox{1.5ex}{${\scriptscriptstyle(\!\frac{\,{u_{l-2}}\,}{v_{l-2}}\!)}$}\hspace{.2em}
         X_{i_{l-1}}\mbox{$-\!\!\!-\!\!\!\longrightarrow$}\hspace{-2.5em}
         \raisebox{1.5ex}{${\scriptscriptstyle (\!\frac{\,{u_{l-1}}\,}{v_{l-1}}\!)}$}\hspace{.2em}
         X_{i_l}.
      \end{equation}
\item The path\eqnref{e-path} is called a {\em cycle around\/}~$X_{i_0}$ if $X_{i_0}=X_{i_l}$.
\item The {\em weight\/} of the path\eqnref{e-path} is defined as
      $\sum_{i=0}^{l-1}\wt(v_i)$.
\item Let $u,\,x,\,v:=uG$ be as in Theorem~\ref{T-ABCD}. If
      $v\in\cC\backslash\cCpol$ we call the infinite path
      \[
        0=x_0\edge{u_0}{v_0}x_1\edge{u_1}{v_1}x_2\edge{u_2}{v_2}x_3\;\cdots\cdots
      \]
      the path associated with the codeword $v=uG$.
      In case $v\in\cCpol$ and $\deg v=N$, we have $x_{N+1}=0$ (since $\deg x\leq\deg v$)
      and we call the finite path
      \[
        0=x_0\edge{u_0}{v_0}x_1\edge{u_1}{v_1}x_2\edge{u_2}{v_2}x_3\;\cdots\cdots\; 
	x_N\edge{u_N}{v_N}x_{N+1}=0
      \]
      the cycle around zero associated with the codeword~$v=uG$.
\end{alphalist}
\end{defi}
Note that the edges of the state diagram correspond to the transitions in
the canonical shift register:
$X_i\edge{u}{v}X_j$ is an edge if for some time instance~$t$ the memory vector is
given by~$X_i$, the input is~$u$, and this leads to the next memory vector~$X_j$ and
the output~$v$.
The only exception is the state $X_0=0$ together with the input~$u=0$. 
This leads to the next state $X_j=X_0A+uB=0$ and $v=X_0C+uD=0$ and this transition 
(or the edge $0\edge{0}{0}0$) is not included in the state diagram.
Hence there emerge $q^k$ edges at each vertex except for the vertex $X_0=0$ at which 
only $q^k-1$ edges emerge.

As a consequence, the state diagram contains all information about the encoding
process.
The cycles around $X_0=0$ correspond to the molecular codewords
in $\im G$
(notice that the edge $0\edge{0}{0}0$ corresponds to the situation $x_L=0$ and
$u_L=0$ occurring only for concatenated, but not tightly concatenated, codewords).
The message sequence determines the path through the graph and the corresponding
$v$-labels yield the associated codeword.
Note also that it is possible to have two different edges between the same vertices, 
i.~e.\ edges of the form $X_i\edge{u}{v}X_j$ and $X_i\edge{u'}{v'}X_j$
where $u\not=u'$.
This happens if and only if $\gamma_l=0$ for at least one~$l$ as can easily be
seen from the matrix~$B$.
In this and only this case there are also edges of the form $0\edge{u}{v}0$ in
the state diagram such that $u\in\ker B$ is not zero.

Since the state space of Example~\ref{E-CCF} has $16^3$ elements, we are not able to 
explicitly display the associated state diagram. 
We will rather restrict ourselves to the following smaller encoder.

\begin{exa}\label{E-213}
Let $G=(1+z+z^2+z^3,\,1+z^2+z^3)\in\F_2[z]^{1\times 2}$.
Thus, $n=2,\,k=1$, and $\gamma=\gamma_1=3$.
The state diagram has $s=2^3=8$ vertices.
We obtain the matrices
\[
   A=\begin{pmatrix}0&1&0\\0&0&1\\0&0&0\end{pmatrix},\
   B=\begin{pmatrix}1&0&0\end{pmatrix},\
   C=\begin{pmatrix}1&0\\1&1\\1&1\end{pmatrix},\
   D=\begin{pmatrix}1&1\end{pmatrix}.
\]
Going through all options for the equations $x_{t+1}=x_tA+u_tB,\,v_t=x_tC+u_tD$ 
yields the state diagram
\\
\mbox{}\hspace*{2.3cm}
\includegraphics[height=5cm]{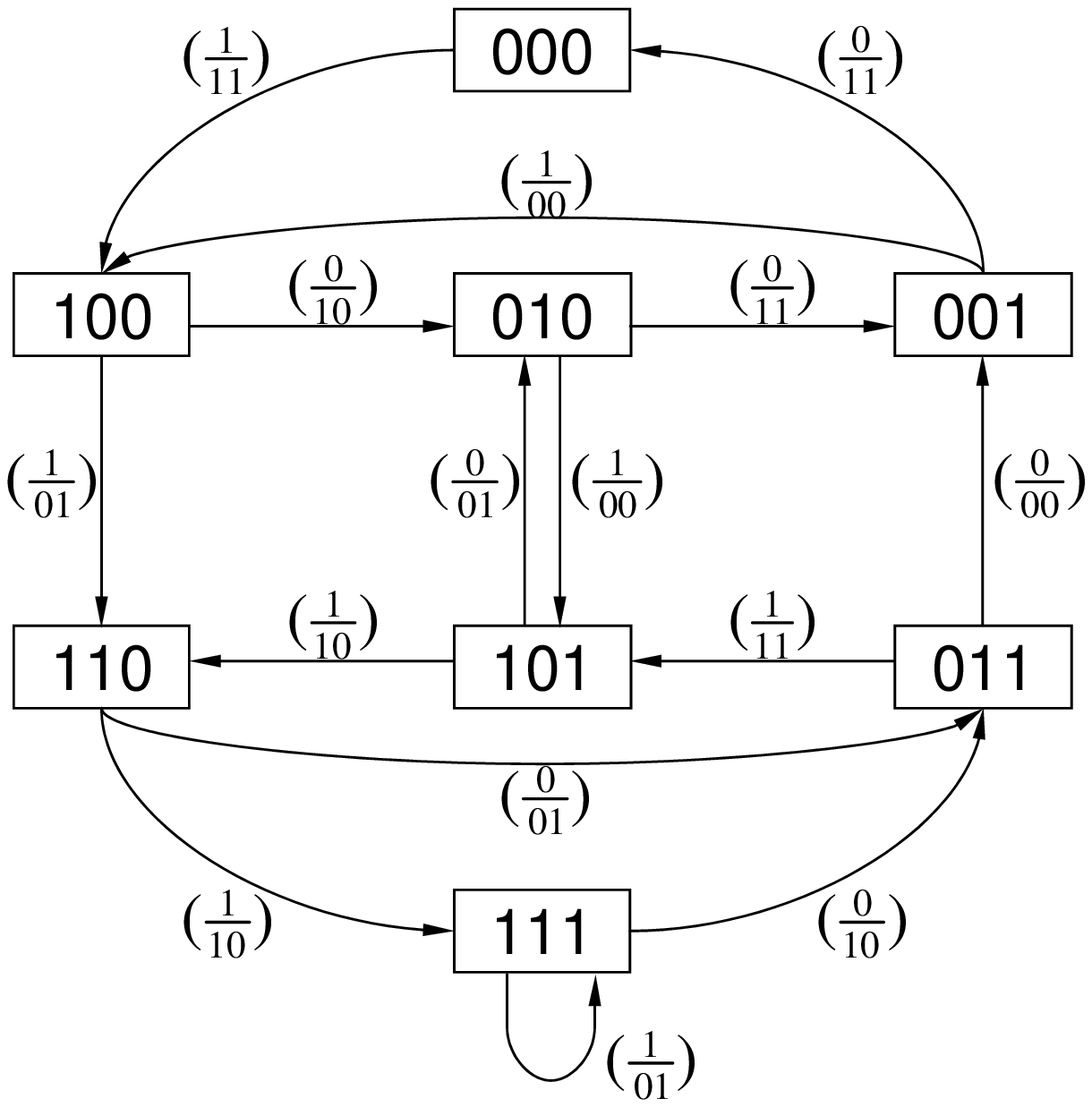}
\end{exa}

The following lemma will be useful for the investigation of cycles of weight zero below.
\begin{lemma}\label{L-zerocycle}
There does not exist a cycle of the form
\[      
   X_{i_0}\edge{0}{0}X_{i_1}\edge{0}{0}X_{i_2}\edge{0}{0}\ldots\ldots\
   \edge{0}{0}X_{i_l}=X_{i_0}
\]
in the state diagram.
\end{lemma}
\begin{proof}
Suppose we have such a cycle. 
Then $X_{i_1}=X_{i_0}A,\,X_{i_2}=X_{i_0}A^2$ and eventually $X_{i_0}=X_{i_0}A^l$.
Running possibly several times through this cycle we may assume without restriction 
that~$l$ is bigger than the nilpotency index of~$A$. Thus $A^l=0$ and hence $X_{i_0}=0$.
But then also $X_{i_1}=0$, which is a contradiction since the edge $0\edge{0}{0}0$ does not 
belong to the state diagram.
\end{proof}

It is possible to characterize basicness, thus non-catastrophicity, of the 
encoder matrix in terms of the state diagram.
As opposed to the main line of this section we do not make any assumptions on the matrix~$G$
in the following proposition.
The controller canonical form and the state diagram can be defined exactly the same way as for basic
polynomial matrices.
Recall from matrix theory that a matrix $G\in\F[z]^{k\times n}$ is basic if and only 
if $\rank G(\lambda)=k$ for all $\lambda$ in an algebraic closure of~$\F$.

\begin{prop}\label{P-basic}
Let $\overline{\F}$ be an algebraic closure of~$\F$ and let
$G\in\F[z]^{k\times n}$ have rank~$k$. 
Consider the associated state diagram.
Then
\begin{arabiclist}
\item $\rank G(0)<k\Longleftrightarrow$ there exists an edge $0\edge{u}{0}X$ for some
      $u\in\F^k\backslash\{0\}$ and $X\in\F^{\gamma}$.
\item $\rank G(\lambda)<k$ for some
      $\lambda\in\overline{\F}\backslash\{0\}\Longleftrightarrow$ there exists a
      cycle of weight zero.
\end{arabiclist}
\end{prop}
\begin{proof}
(1) is clear since $G(0)=D$.
\\
(2) ``$\Rightarrow$'':
It is well-known that the assumption implies the existence of
some $u=\sum_{t\geq0}u_tz^t\in\Fpower^k\backslash\F[z]^k$ such that
$v=uG=\sum_{t=0}^Lv_tz^t\in\F[z]^n$, hence $v_t=0$ for $t>L$, see 
\cite[Thm.~6.3]{McE98}. 
Let $x=uB(z^{-1}I-A)^{-1}=\sum_{t\geq0}x_tz^t$ be the corresponding state 
sequence.
Hence each~$x_t$ is a vertex of the state diagram.
Since the state diagram has only finitely many edges, but the information sequence has 
infinitely many nonzero blocks~$u_t$, there have to be some $t_2>t_1>L$ and a vertex~$X$ 
such that $x_{t_1}=X=x_{t_2}$.
Since $v_t=0$ for $t>L$ this gives us a cycle of weight zero.
\\
 ``$\Leftarrow$'':
Suppose we have a cycle of weight zero and length~$l$ in the state diagram, say
\[
   X_{i_0}\edge{\hat{u}_0}{0}X_{i_1}\edge{\hat{u}_1}{0}X_{i_2}\edge{\hat{u}_2}{0}\ldots\ldots\
   \mbox{$-\!\!\!-\!\!\!\longrightarrow$}\hspace{-2.5em}
   \raisebox{1.5ex}{${\scriptscriptstyle(\!\frac{\,{\hat{u}_{l-2}}\,}{0}\!)}$}\hspace{.2em}
   X_{i_{l-1}}\mbox{$-\!\!\!-\!\!\!\longrightarrow$}\hspace{-2.5em}
   \raisebox{1.5ex}{${\scriptscriptstyle (\!\frac{\,{\hat{u}_{l-1}}\,}{0}\!)}$}\hspace{.2em}
   X_{i_l}
\]
where $X_{i_0}=X_{i_l}=:X$.
By Lemma~\ref{L-zerocycle} not all $\hat{u}_i$ are zero.
There exists a path from the state $X_0=0$ to the state~$X$, say
\[
   0\edge{u_0}{v_0}X_{j_1}\edge{u_1}{v_1}X_{j_2}\edge{u_2}{v_2}\ldots\ldots\
   \mbox{$-\!\!\!-\!\!\!\longrightarrow$}\hspace{-2.5em}
   \raisebox{1.5ex}{${\scriptscriptstyle (\!\frac{\,{u_{T-1}}\,}{v_{T-1}}\!)}$}\hspace{.2em}
   X_{j_T}=X.
\]
This can be easily seen since the existence of such a path is equivalent to the existence
of $(u_0,\ldots,u_{T-1})$ such that 
\[
  X=(u_0,\ldots,u_{T-1})\begin{pmatrix}BA^{T-1}\\ BA^{T-2}\\ \vdots\\ B\end{pmatrix},
\]
which in turn follows from the full column rank of the right hand side matrix if we
choose~$T$ big enough.
Put $\hat{u}:=\sum_{t=0}^{l-1}\hat{u}_tz^t$ and
$u=\sum_{t=0}^{T-1}u_tz^t+\sum_{t=0}^{\infty}\hat{u}z^{T+tl}\in\Fpower^k$.
Then $u\not\in\F[z]^k$, since $\hat{u}\not=0$, 
but $v=uG=\sum_{t=0}^{T-1}v_tz^t\in\F[z]^n$.
By \cite[Thm.~6.3]{McE98} this implies the left hand side of~(2).
\end{proof}

\section{The adjacency matrix and the weight distribution}\label{S-3}
\setcounter{equation}{0}
In this section we will show how to compute the weight distribution of a convolutional
code in terms of the state diagram.
A main tool will be the adjacency matrix associated with the state diagram as it has been 
defined in~\cite{McE98a}.
In slightly different forms this matrix appears also in other papers on convolutional 
codes, see for instance \cite[Sec.~3.10]{JoZi99}. 
Theorem~\ref{T-weightdistr} has been derived in~\cite{McE98a}.
First of all we need a reasonable definition for the weight distribution.
Of course, it is sufficient to count only the weights of atomic codewords.
In order to do so, we need to show that we are dealing with finite numbers if
counted appropriately. This will be dealt with in the first lemma.

\begin{ass}\label{GA}
From now on we will always assume that the generator matrix
$G\in\F[z]^{k \times n}$ of the code~$\cC\subseteq\Flaurent^n$ is 
minimal with Forney indices $\gamma_1,\ldots,\gamma_k$ and overall constraint 
length~$\gamma>0$.
Moreover, we define the matrices $A,\,B,\,C$, and~$D$ as in Definition~\ref{D-ABCD}.
Finally, let $s:=q^\gamma$ and $\F^{\gamma}=\{X_0=0,X_1,\ldots,X_{s-1}\}$ be the vertices 
of the state diagram in some fixed ordering.
\end{ass}

\begin{lemma}\label{L-atomic}
Assume the memory of~$G$, i.~e., the maximal Forney index, is given by~$m$.
\begin{alphalist}
\item Let $v=uG\in\cCpol$ where $u=\sum_{i=0}^Lu_iz^i\in\F[z]^k$ with $u_0\not=0\not=u_L$.
      If $u_{l+1}=u_{l+2}=\ldots=u_{l+m}=0$ for some $l\in\{0,\ldots,L-m-1\}$,
      then~$v$ is concatenated.
\item For all $\alpha\in\N_0$ we have
      $\#\{v\in\cCpol\mid  v\text{ atomic},\;\wt(v)\leq\alpha\}<\infty$.
\end{alphalist}
\end{lemma}
\begin{proof}
(a) Write $u=\sum_{i=0}^lu_iz^i+\sum_{i=l+m+1}^Lu_iz^i=:\tilde{u}+\hat{u}$.
Then $\deg(\tilde{u}G)\leq l+m$ and $\hat{u}G\in z^{l+m+1}\F[z]^n$, thus
$v=\tilde{u}G+\hat{u}G$ is concatenated at some time $t\leq l+m+1$.
\\
(b) Let $\hat{G}\in\F[z]^{n\times k}$ be a right inverse of~$G$ and let~$\hat{G}$ have
maximal row degree~$\hat{m}$.
Suppose $v=uG\in\cCpol$ is a codeword with at least $m+\hat{m}$ consecutive zero coefficients. 
Thus, $v=\tilde{v}+\hat{v}$ where 
$\tilde{v},\,\hat{v}\in\F[z]^n\backslash\{0\}$ satisfy $\deg\tilde{v}\leq L$ and 
$\hat{v}\in z^{L+m+\hat{m}+1}\F[z]^n$ for some $L\in\N_0$.
Then $u=uG\hat{G}=\tilde{v}\hat{G}+\hat{v}\hat{G}$ and part~(a) shows that~$v$ is not 
atomic.
All this proves that atomic codewords do not have more than $m+\hat{m}-1$ consecutive zero coefficients. 
As a consequence, all atomic codewords of weight at most~$\alpha$ have a degree bounded 
by some $M_{\alpha}\in\N$ proving the assertion.
\end{proof}

\begin{rem}\label{R-concatu}
It is not hard to prove that if all Forney indices of~$G$ are equal to~$m$ then 
part~(a) above becomes an if-and-only-if statement. 
Indeed, let~$v=uG\in\cCpol$ be concatenated as $v=\tilde{v}+\hat{v}$ where 
$\tilde{v},\,\hat{v}\in\cCpol\backslash\{0\}$ and 
$\deg\tilde{v}\leq T$ and $\hat{v}\in z^{T+1}\F[z]^n$ for some $T\leq\deg u-1$.
We have to show that $u_{T}=u_{T-1}=\ldots=u_{T-m+1}=0$.
From Lemma~\ref{L-concatenated} we know that $x_{T+1}=0$, thus we obtain recursively
\begin{equation}\label{e-xT}
 \begin{array}{rcl}
  0&=&x_{T}A+u_TB=x_{T-1}A^2+u_{T-1}BA+u_TB=\ldots\\
   &=&x_{T-m+1}A^m+(u_{T-m+1},\ldots,u_T)
     \begin{pmatrix}BA^{m-1}\\ \vdots\\BA\\B\end{pmatrix}.
 \end{array}
\end{equation}
It follows directly from the definition of the controller canonical form that if all 
Forney indices are~$m$, then $A^m=0$ and the matrix on the very right 
of\eqnref{e-xT} is a non-singular $\gamma\times\gamma$-matrix.
Hence\eqnref{e-xT} yields the desired result.
\end{rem}
The above makes the computation of the distance of a given code to a finite problem, at least 
theoretically.
\begin{defi}\label{D-weightdistr}
For all $\alpha,\,l\in\N$ define
\[
  \omega_{l,\alpha}:=\#\{v\in\cCpol\mid v\text{ atomic}, \deg v=l-1,\,
                            \wt(v)=\alpha\}.
\]
The power series
${\DS \Omega=\Omega(W,L):=\sum_{l=1}^{\infty}\sum_{\alpha=1}^{\infty}
          \omega_{l,\alpha}W^{\alpha}L^l\in\Q[\![W,L]\!]}$
is called the {\em weight distribution\/} of the code~$\cC$.
\end{defi}
Observe that the weight distribution~$\Omega$ is an invariant of the code and does 
not depend on a chosen generator matrix.
The definition above is somewhat unusual as it omits the constant term $1$
representing the zero codeword.
We will see later that our definition is more convenient.
Note that the numbers $\omega_{l,\alpha}$ are indeed finite.
Moreover, since each codeword of degree~$l-1$ has at most~$l$ nonzero coefficients
in~$\F^n$, we have 
$\sum_{\alpha=1}^{\infty}\omega_{l,\alpha}W^{\alpha}=\sum_{\alpha=1}^{nl}\omega_{l,\alpha}W^{\alpha}$, 
thus
\begin{equation}\label{e-weight1}
  \Omega=\sum_{l=1}^{\infty}\sum_{\alpha=1}^{nl}
          \omega_{l,\alpha}W^{\alpha}L^l\in\Q[W][\![L]\!].
\end{equation}
Notice also that for block codes we simply have 
$\Omega=\sum_{\alpha=1}^n\omega_{1,\alpha}W^{\alpha}L$. 
This is, up to the factor $L$, the ordinary weight distribution for block codes where the 
constant term~$1$, representing the zero codeword, has been omitted. 
In the general case we also have for each $\alpha\in\N$ that
$\sum_{l=0}^{\infty}\omega_{l,\alpha}L^l$ is a finite sum, due to
Lemma~\ref{L-atomic}(b). Hence
\begin{equation}\label{e-weight2}
    \Omega=\sum_{\alpha=1}^{\infty}\sum_{l=0}^{\infty}\omega_{l,\alpha}L^lW^{\alpha}\in\Q[L][\![W]\!].
\end{equation}
Finally observe that
\[
   \dist(\cC)=\min\Big\{\alpha\in\N\,\Big|\,
   \exists\,l\in\N:\;\omega_{l,\alpha}\not=0\Big\}
\]
is the degree of the smallest term ocurring in the series expansion with respect to~$W$.

In the sequel we will show how one can compute, at least theoretically, the weight 
distribution of a code using the state diagram.
All necessary information is contained in the following matrix~$\Lambda$.
Recall General Assumption~\ref{GA}.
For ease of notation rows and columns of $s\times s$-matrices will always be indexed 
with $i,j=0,\ldots,s-1$.
\begin{defi}\label{D-adjacency}
For all $(i,j)\in\{0,\ldots,s-1\}^2$ and all $\alpha\in\{0,\ldots,n\}$ define
\[
  \lambda_{i,j}^{(\alpha)}:=\left\{
    \begin{array}{cl}
       0,&\text{if } (i,j,\alpha)=(0,0,0),\\[1ex]
       \#\{u\in\F^k\mid X_j=X_iA+uB,\,\wt(X_iC+uD)=\alpha\},& \text{else}.
    \end{array}\right.
\]
Furthermore, define the matrix
\[
   \Lambda:=\left(\sum_{\alpha=0}^n\lambda_{i,j}^{(\alpha)}W^{\alpha}\right)_{i,j=0,\ldots,s-1}
   \in\Q[W]^{s\times s}.
\]
We call~$\Lambda$ the {\em adjacency matrix\/} of the encoder matrix~$G$ (or of the state
diagram).
\end{defi}
Notice that $\lambda_{i,j}^{(\alpha)}$ is the number of all edges in the state diagram of the 
form $X_i\edge{u}{v}X_j$ where $\wt(v)=\alpha$.
Obviously the weight is bounded by~$n$.
The identity $\lambda_{0,0}^{(0)}=0$ reflects the fact that there is no edge of the form 
$0\edge{u}{0}0$ in the state diagram.
One should also observe that the adjacency matrix depends on the choice of the encoder~$G$,
thus on the matrices~$A,B,C$, and~$D$ as well as on the numbering of the states.
We will discuss this issue in Section~\ref{S-AdjUnique} in detail.

\begin{exa}\label{E-Lambda}
Let $\F=\F_2$.
\begin{arabiclist}
\item Let $G=\begin{pmatrix}1&1&0\\0&z+1&z\end{pmatrix}\in\F[z]^{2\times3}$.
      Thus $\gamma=1$ and $s=2$ and
      $A=\begin{pmatrix}0\end{pmatrix},\ B=\begin{pmatrix}0\\1 \end{pmatrix},\
        C=\begin{pmatrix}0&1&1\end{pmatrix},\
        D=\begin{pmatrix}1&1&0\\0&1&0\end{pmatrix}$.
      With this one obtains straightforwardly
      $\Lambda=\begin{pmatrix}W^2&2W\\2W^2&W+W^3\end{pmatrix}$.
\item In Example~\ref{E-213} we have $s=8$ and obtain, with an appropriate ordering 
      of the states,
      \[
         \Lambda=\begin{pmatrix}
             0&0&0&0&W^2&0&0&0\\W^2&0&0&0&1&0&0&0\\
             0&W^2&0&0&0&1&0&0\\0&1&0&0&0&W^2&0&0\\
             0&0&W&0&0&0&W&0\\0&0&W&0&0&0&W&0\\
             0&0&0&W&0&0&0&W\\ 0&0&0&W&0&0&0&W\end{pmatrix}.
      \]
\end{arabiclist}
\end{exa}

The following proposition will lead us to a computation of the weight distribution
via the adjacency matrix~$\Lambda$.

\begin{prop}\label{P-Lambda1}
Let $l\in\N$. For $(i,j)\in\{0,\ldots,s-1\}^2$ denote the entry at the position $(i,j)$
of $\Lambda^l$ by
\[
  (\Lambda^l)_{i,j}=\sum_{\alpha=0}^{nl}\lambda_{i,j}^{(l,\alpha)}W^\alpha\in\Q[W].
\]
Then $\lambda_{i,j}^{(l,\alpha)}$ is the number of paths in the state diagram
from~$X_i$ to~$X_j$ of length~$l$ and weight~$\alpha$.
In particular, $\lambda_{0,0}^{(l,0)}=0$ for all $l\in\N$.
\end{prop}
Note that it is clear that the entries in $\Lambda^l$ are all polynomials of
degree at most $nl$.
\\[1ex]
\begin{proof}
We proceed by induction on~$l$. For $l=1$ the assertion is just the definition of
the adjacency matrix~$\Lambda$.
Assume the assertion is true for $l\geq1$.
Then
\begin{align*}
  (\Lambda^{l+1})_{i,j}&=\sum_{\nu=0}^{s-1}\Lambda_{i,\nu}(\Lambda^l)_{\nu, j}
       =\sum_{\nu=0}^{s-1}\Big(\sum_{\alpha=0}^n\lambda_{i,\nu}^{(1,\alpha)}W^\alpha\Big)
           \Big(\sum_{\alpha=0}^{nl}\lambda_{\nu, j}^{(l,\alpha)}W^\alpha\Big)\\
       &=\sum_{\alpha=0}^{n(l+1)}\sum_{\nu=0}^{s-1}\sum_{\beta=0}^{\alpha}\lambda_{i,\nu}^{(1,\beta)}
             \lambda_{\nu ,j}^{(l,\alpha-\beta)}W^{\alpha}
\end{align*}
By induction hypothesis $\lambda_{\nu, j}^{(l,\alpha-\beta)}$ is the number of paths
from~$X_{\nu}$ to~$X_j$ of length~$l$ and weight $\alpha-\beta$ and likewise
$\lambda_{i,\nu}^{(1,\beta)}$ is the number of edges from~$X_i$ to~$X_{\nu}$ and
weight~$\beta$.
Since all paths from~$X_i$ to~$X_j$ go through exactly one state~$X_{\nu}$ after one
step, we obtain that the total number of paths from $X_i$ to~$X_j$ of length~$l+1$
and weight~$\alpha$ is given by
$\sum_{\nu=0}^{s-1}\sum_{\beta=0}^{\alpha}\lambda_{i,\nu}^{(1,\beta)}
 \lambda_{\nu, j}^{(l,\alpha-\beta)}$.
This proves the first assertion.
The second one is clear since there are no cycles of weight zero in the state diagram according to
Proposition~\ref{P-basic}(2).
\end{proof}

Now we are in a position to present the main result about the weight distribution.
Recall that only atomic codewords are counted.
According to Lemma~\ref{L-concatenated} they correspond to cycles around the state $X_0=0$ 
in the state diagram that do not pass through this state in the meantime.
The number $\lambda_{0,0}^{(l,\alpha)}$ counts all cycles around $X_0=0$, but does not take 
into account whether it passes through this state in the meantime.
Therefore, $\lambda_{0,0}^{(l,\alpha)}$ is the number of {\em molecular\/} codewords of 
length~$l$ and weight~$\alpha$ (this will also become clear from the proof below).
The next theorem shows how to obtain from this the weight distribution~$\Omega$.
For the sake of completeness we also add the proof as it can be found in~\cite{McE98a}.

\begin{theo}[McEliece~\mbox{\cite[Thm.~3.1]{McE98a}}]\label{T-weightdistr}
Let
$\Phi:=1+\sum_{l=1}^{\infty}\sum_{\alpha=1}^{\infty}\lambda_{0,0}^{(l,\alpha)}W^\alpha L^l
 \in\Q[W][\![L]\!]$. 
Then
\begin{alphalist}
\item $\Phi=\big((I-L\Lambda)^{-1}\big)_{0,0}\in\Q(W,L)$, the entry in the upper left
      corner of the rational matrix~$(I-L\Lambda)^{-1}\in\Q(W,L)^{s\times s}$.
\item $\Omega=1-\Phi^{-1}$.
\end{alphalist}
\end{theo}
This is also compliant with the block code case in which 
$\Lambda=\sum_{v\in\cC\backslash\{0\}}W^{\text{wt}(v)}\in\Q[W]$ and $\Omega=\Lambda L$.
\\[1ex]
\begin{proof}
(a) is seen via
\[
  \Phi=1+\sum_{l=1}^{\infty}(\Lambda^l)_{0,0}L^l
   =\Big(I+\sum_{l=1}^{\infty}(\Lambda L)^l\Big)_{0,0}
   =\Big(\sum_{l=0}^{\infty}(\Lambda L)^l\Big)_{0,0}
   =\Big((I-\Lambda L)^{-1}\Big)_{0,0}.
\]
(b) Recall from Equation\eqnref{e-weight1} that
$\Omega=\sum_{l=1}^{\infty}\sum_{\alpha=1}^{nl}\omega_{l,\alpha}W^{\alpha}L^l$.
We first show that for $r>1$ the coefficient of~$W^\alpha L^l$ in the power $\Omega^r$ is 
the number of all codewords of weight~$\alpha$ and degree~$l-1$ that consist of~$r$ 
tightly concatenated atomic codewords.
In other words, it is the number of all cycles around $X_0=0$ in the state diagram that 
have length~$l$ and weight~$\alpha$, and that pass through the zero state exactly $r-1$ 
times except for start and endpoint. 
This can be proven by induction on~$r$.
Precisely, let
$\Omega^r=\sum_{l=1}^\infty\sum_{\alpha=1}^\infty\omega_{l,\alpha}^{(r)}W^\alpha L^l$, then
\[
  \omega_{l,\alpha}^{(r+1)}=\sum_{l'=1}^{l-1}\sum_{\beta=1}^{\alpha-1}
  \omega_{l',\beta}\omega_{l-l',\alpha-\beta}^{(r)}
\]
and $\omega_{l',\beta}$ is the number of all atomic cycles around $X_0=0$ of length~$l'$ and weight~$\beta$
while $\omega_{l-l',\alpha-\beta}^{(r)}$ is, by induction, the number of all cycles around
$X_0=0$ of length~$l-l'$ and weight~$\alpha-\beta$ that pass exactly $r-1$ through
$X_0=0$ except for start and endpoint.
Since each cycle consisting of~$r$ tightly concatenated atomic cycles can be
obtained in a unique way by tightly concatenating these types of cycles we obtain the desired
assertion about the coefficients of~$\Omega^r$.
\\
Using Proposition~\ref{P-Lambda1} we conclude
\[
   \Phi=1+\Omega+\Omega^2+\Omega^3+\ldots=\sum_{r=0}^\infty \Omega^r=\frac{1}{1-\Omega}
\]
and this yields $\Omega=1-\Phi^{-1}$.
\end{proof}
Observe that the proof shows indeed that $\Phi$ is the weight 
distribution of the molecular codewords.

If one follows the arguments above one has to perform the following steps in 
order to compute the weight distribution of a given code:
\begin{arabiclist}
\item Compute $\Lambda$.
\item Solve the equation $(I-L\Lambda)x=(1,0,\ldots,0)\T$ for $x\in\Q(W,L)^s$.
\item Then $\Phi=x_1$ and $\Omega=1-\Phi^{-1}$.
\end{arabiclist}
While~(1) and~(3) are easily done for reasonably sized parameters, step~(2) quickly becomes 
unpractical with growing overall constraint length~$\gamma$ and/or field size~$q$ since 
the size of the adjacency matrix is $q^{\gamma}\times q^{\gamma}$. 
Better algorithms for computing the weight distribution while avoiding the big 
adjacency matrix can be found in \cite{FiNo91,MoHe99,On90}.

\begin{exa}\label{E-WeightDistr}
Consider $\F=\F_2$ and $G:=\begin{pmatrix}1+z+z^2+z^3&1+z^2+z^3\end{pmatrix}$.
The matrix is basic and, of course, minimal.
In this case the adjacency matrix is only $8\times8$ and we can perform the computation
of the weight distribution along the steps~(1) --~(3) above by using, for instance, 
Maple.
We computed the adjacency matrix~$\Lambda$ already in Example~\ref{E-Lambda}(2). 
From that one obtains the weight distribution
\begin{align*}
  \Omega&=L^4W^6(L\!+\!W\!-\!LW^2)/(1\!-\!LW\!-\!L^2W\!+\!L^3W^2\!-\!L^3W^3\!
                     -\!L^4W^2\!-\!L^3W^4\!+\!L^4W^4)\\
   &=L^5W^6\!+\!(L^4\!+\!L^6\!+\!L^7)W^7\!+\!(L^6\!+\!L^7\!+\!L^8\!+\!2L^9)W^8\!+\!
     (4L^8\!+\!L^9\!+\!3L^{10}\!+\!3L^{11})W^9\\
   &\quad \!+\!O(W^{10}).
\end{align*}
Thus, the distance is~$6$ and the code contains exactly one atomic codeword of
weight~$6$, it has length~$5$.
It contains three atomic codewords (of length~$4,\,6$, and~$7$, respectively) of weight~$7$
and~$5$ atomic codewords of weight~$8$, two of which have length~$9$ the other
three have length~$6,\,7,\,8$, respectively, etc.
\end{exa}

At the end of this section we want to briefly mention two distance parameters appearing 
in a totally different context in the literature that are closely related to our notions.
In \cite[p.~541]{JPB90} the {\em extended row distances\/} $\hat{d}^r_l$ are defined. 
The definition shows immediately that 
\[
   \hat{d}^r_l=\min\{\wt(v)\mid v\text{ atomic}, \deg(v)=l\}
              =\min\{\alpha\in\N_0\mid \omega_{l+1,\alpha}\not=0\}
\]
where $\omega_{j,\alpha}$ are given in Definition~\ref{D-weightdistr}.
Thus these distance parameters can be recovered, at least theoretically, from the weight 
distribution~$\Omega$.
There are also other definitions of the extended row distances in the literature.
They are slightly different and in general not that closely related to atomic or 
molecular codewords and some of them even depend on the choice of the (minimal) encoder. 

In \cite[p.~155]{HJZ02} the $l$th order {\em active burst distance\/} of the code $\cC=\im G$
is defined as 
\[
  a^b_l:=\min\{\wt\big((uG)_{[0,l]}\big)\mid x_{l+1}=0\text{ and }
            (x_i,x_{i+1},u_i)\not=0\text{ for all }0\leq i\leq l\},
\]
where, as usual,~$x$ is the associated state sequence, see also \cite[Sec.~3.2]{JoZi99}.
Moreover, $(uG)_{[0,l]}$ denotes the codeword truncated after the $l$th power of~$z$.
As we will show now, if~$G$ is minimal then
\begin{equation}\label{e-abd}
  a^b_l=\min\{\wt(v)\mid v\text{ molecular}, \deg v=l\}
     =\min\{\alpha\in\N\mid \lambda_{0,0}^{(l+1,\alpha)}\not=0\}
\end{equation}
where $\lambda_{i,j}^{(l,\alpha)}$ are defined as in Proposition~\ref{P-Lambda1}.
The second identity follows directly from the discussion right before 
Theorem~\ref{T-weightdistr}.
As for the first identity, note first that the condition $(x_i,x_{i+1},u_i)\not=0$
simply means that the codeword $uG$ is molecular.
Moreover, Lemma~\ref{L-concatenated} along with $x_{l+1}=0$ implies that 
$v:=(uG)_{[0,l]}$ is a codeword. 
Finally we have $\deg v=l$, which can be seen as follows.
Suppose $\deg v<l$. Since~$G$ is minimal we have $\deg u\leq\deg v$, see 
Theorem~\ref{T-ABCD}(2).
Hence $u_l=0=v_l$. 
Using the controller canonical form $(A,B,C,D)$ we obtain $x_{l+1}=0=x_lA$ and 
$v_l=0=x_lC$, and Lemma~\ref{L-ABCD}(2) implies $x_l=0$.
Hence $(x_l,x_{l+1},u_l)=0$ which contradicts the choice of~$v$.
All this together shows that~$v$ is a molecular codeword of degree~$l$.
Conversely one can easily see that each such codeword has a state sequence~$x$ such 
that $(x_i,x_{i+1},u_i)\not=0$ for all $0\leq i\leq l$.
This proves the first identity of\eqnref{e-abd}.
Hence the active burst distances occur in the series~$\Phi$ as used in
Theorem~\ref{T-weightdistr} for enumerating the molecular codewords.
In particular we have that $\min_{l\geq0}a^b_l$ equals the free distance of the code,
see also \cite[Thm.~3.8]{JoZi99}.

\section{The adjacency matrix as an invariant of the code}\label{S-AdjUnique}
\setcounter{equation}{0}
In this section we will prove that the adjacency matrix is an invariant of the code, i.~e., 
does not depend on the choice of the minimal generator matrix, provided that one factors out 
the effect of the (arbitrarily chosen) ordering of the states.
As to our knowledge the result of this and the next section are new.

Throughout this section let the data be as in General Assumption~\ref{GA}.
It is clear from Definition~\ref{D-adjacency} that the adjacency matrix of~$G$ 
depends on the ordering of the states. 
Suppose now we have fixed two different orderings on the state space $\F^{\gamma}$, each 
one satisfying $X_0=0$. 
The we obtain two (different) adjacency matrices~$\Lambda_G$ and~$\Lambda_G'$ and it
is clear that
\begin{equation}\label{e-Pi}
   \Lambda_G'=\Pi\Lambda_G\Pi^{-1}
   \text{ for some permutation matrix $\Pi\in Gl_{s}(\Q)$ such that }
   \Pi_{0,0}=1.
\end{equation}
Again, we will always index the rows and columns of matrices in 
$\Q[W]^{s\times s}$ by $i,\,j=0,\ldots,s-1$.
We define
\[
  \cG:=\{\Pi\mid \Pi\in Gl_{s}(\Q)\text{ is a permutation matrix and } \Pi_{0,0}=1\}
\]
to be the group of these specific permutation matrices and the action
\[
  \cG\times\Q[W]^{s\times s}\longrightarrow \Q[W]^{s\times s},\quad
  (\Pi,\Lambda)\longmapsto \Pi\Lambda\Pi^{-1}.
\]
This yields a group action on $\Q[W]^{s\times s}$ and we obtain the quotient space
$\Q[W]^{s\times s}/_{\textstyle\cG}$ of all equivalence classes
\[
   \overline{\Lambda}:=\{\Pi\Lambda\Pi^{-1}\mid \Pi\in\cG\},\ 
   \Lambda\in\Q[W]^{s\times s}.
\]
All this gives us a well-defined mapping $G\longmapsto \overline{\Lambda_G}$
from the set of all minimal generator matrices with overall constraint length~$\gamma$
into $\Q[W]^{s\times s}/_{\textstyle\cG}$ by simply choosing $\Lambda_G$ as the adjacency matrix of~$G$ 
with respect to any arbitrary ordering of the state space such that $X_0=0$.
We will show now that $\overline{\Lambda_G}$ is even an invariant of the code.
Indeed, we have
\begin{theo}\label{T-uniqueAdj}
Let $G,\,G'\in\F[z]^{k\times n}$ be two minimal generator matrices such that
$\cC:=\im G=\im G'$. Then 
\begin{equation}\label{e-Lambda}
  \overline{\Lambda_G}=\overline{\Lambda_{G'}}.
\end{equation}
Thus, the adjacency matrices of~$G$ and~$G'$ differ only via conjugation by some matrix
in~$\cG$.
We will write $\bar{\Lambda}(\cC):=\overline{\Lambda_G}$ for this invariant and call
it the {\em generalized adjacency matrix of the code}.
\end{theo}
The theorem tells us that the adjacency matrix is, up to the group action of~$\cG$, 
an invariant of the code.
In other words, the generalized adjacency matrix is a well-defined mapping
\begin{equation}\label{e-Cadjac}
\begin{split}
  \bar{\Lambda}:\;\{\cC\subseteq\F_q(\!(z)\!)^n\mid\cC\text{ code with overall constraint length }\gamma\}
  &\longrightarrow \Q[W]^{q^{\gamma}\times q^{\gamma}}/_{\textstyle\cG}\\
  \cC\hspace*{3cm}&\longmapsto \qquad \bar{\Lambda}(\cC)
\end{split}
\end{equation}
\begin{proof}
Let $\nu_1,\ldots,\nu_k$ and $\mu_1,\ldots,\mu_k$ be the row degrees of~$G$ and~$G'$, 
respectively. 
By assumption and\eqnref{e-cpolyunique} we have 
\begin{equation}\label{e-G'G}
  G'=UG\text{ for some matrix } U\in Gl_k(\F[z])
\end{equation}
and we also have $\{\nu_1,\ldots,\nu_k\}=\{\mu_1,\ldots,\mu_k\}$.
Let $\gamma:=\sum_{i=1}^k\nu_i=\sum_{i=1}^k\mu_i$ be the overall constraint length of the code.
\\
Since every unimodular matrix $U\in Gl_k(\F[z])$ is the product of elementary matrices, 
we may show the result for each type of elementary transformation separately. 
In the rest of the proof we fix an ordering on the state space such that $X_0=0$.
Moreover we define $(A,B,C,D)$ and $(A',B',C',D')$ to be the controller canonical 
forms of~$G$ and~$G'$, respectively. 
\\
1) We show that a permutation of the rows of~$G$ results in a conjugation 
of~$\Lambda_G$ just like in\eqnref{e-Pi}.
Thus let us assume $G'=UG$ where~$U$ permutes the $i$th and $j$th row of~$G$.
Then~$A'$ is obtained from~$A$ by permuting the $i$th and $j$th block row and column,
$B'$ is obtained from~$B$ by permuting the $i$th and $j$th row and the $i$th and $j$th
block column, $C'$ is obtained from~$C$ by permuting the $i$th and $j$th block row and 
finally~$D'$ is obtained from~$D$ by permuting the $i$th and $j$th row.
Thus, there exists a permutation matrix $P\in Gl_{\gamma}(\F)$ such that
\[
  A'=PAP^{-1},\ B'=UBP^{-1},\ C'=PC,\ D'=UD 
\]
(this is also correct if~$\nu_i$ or~$\nu_j$ is zero).
Now let $X_l\edge{u}{v}X_m$ be an edge in the state diagram of~$G$, thus 
$(\Lambda_G)_{l,m}=W^{\text{wt}(v)}$. 
Then $X_m=X_lA+uB$ and $v=X_lC+uD$. 
From this we obtain $X_mP^{-1}=(X_lP^{-1})A'+(uU^{-1})B'$ and $v=(X_lP^{-1})C'+(uU^{-1})D'$,
hence 
$X_lP^{-1}\mbox{$-\!\!\!-\!\!\!-\!\!\!\longrightarrow$}\hspace{-3em}
 \raisebox{1.5ex}{${\scriptscriptstyle (\!\frac{\,{uU^{-1}}\,}{v}\!)}$}\hspace{.4em}
X_mP^{-1}$ 
is an edge in the state diagram of~$G'$. 
Since the mapping~$P:\,\F^{\gamma}\rightarrow\F^\gamma$ is simply a permutation of the 
states with $X_0P=X_0$, we arrive at $\Pi\Lambda_G\Pi^{-1}=\Lambda_{G'}$ for some 
suitable permutation $\Pi\in\cG$.
This in turn implies\eqnref{e-Lambda}.
\\
2) Next we consider the case where the matrix~$U$ in\eqnref{e-G'G} multiplies the 
rows of~$G$ with some nonzero constants, say $U=\text{diag}(u_1,\ldots,u_k)\in Gl_k(\F)$.
Then
\[
   A'=A,\, B'=B,\ C'=\hat{U}C,\ D'=UD
\]
where 
\[
   \hat{U}=\begin{pmatrix} u_1I_{\nu_1} &            &   &  \\
                                        &u_2I_{\nu_2}&   &   \\
					&            &\ddots&   \\
					&            &      &u_kI_{\nu_k}\end{pmatrix}.
\]
Here $I_j$ denotes the $j\times j$-identity matrix.
It is easy to see that $\hat{U}A=A\hat{U}$ and $UB=B\hat{U}$, thus 
$A'=A=\hat{U}A\hat{U}^{-1}$ and $B'=B=UB\hat{U}^{-1}$.
Using the same arguments as in case~1) we arrive at $\Pi\Lambda_G\Pi^{-1}=\Lambda_{G'}$ 
for some $\Pi\in\cG$.
\\
3) Now we consider the case where~$U$ adds a {\em constant\/} multiple of one row to another.
Because of part~1) of this proof we may assume $\nu_1=\mu_1\leq\ldots\leq\nu_k=\mu_k$. 
Since we did already part~2) of this proof and since~$G$ and $G'=UG$ are both minimal we 
only have to consider the case where the $j$th row is added to the $i$th row and $j<i$.
Hence $U=I_k+E$ where $E\in\F^{k\times k}$ has a~$1$ at position~$(i,j)$ and~$0$ elsewhere.
Let us first assume $\nu_j>0$.
Put
\begin{equation}\label{e-Uhat}
 \hat{U}=\begin{pmatrix}
            I_{\nu_1}&      &         &      &         &      &  \\
	             &\ddots&         &      &         &      &  \\
		     &      &I_{\nu_j}&      &         &      &  \\
		     &      &         &\ddots&         &      &  \\
		     &      &   M     &      &I_{\nu_i}&      &   \\
		     &      &         &      &         &\ddots&   \\
		     &      &         &      &         &      &I_{\nu_k} 
         \end{pmatrix}
 \in Gl_{\gamma}(\F)
\end{equation}
where 
\[
 M=\begin{pmatrix}I_{\nu_j}\\ 0\end{pmatrix}\in\F^{\nu_i\times\nu_j}.
\]
Then we have
\[
  A'=A,\ B'=B,\ C'=\hat{U}C,\ D'=UD.
\]
Furthermore, it is easy to see that $MA_{j}=A_iM$ and $B_iM=B_j$ where $A_i,\,B_i$ are 
the diagonal blocks of the matrices~$A,\,B$ as given in Definition~\ref{D-ABCD}.
From this we obtain $\hat{U}A=A\hat{U}$ and $UB=B\hat{U}$.
Now we can use the same arguments as in case~2) to finish the proof.
If $\nu_j=0$ we have $A'=A,\,B'=B,\,C'=C$, and $D'=UD$. 
Using $\hat{U}:=I_{\gamma}$ and the fact that the $j$th row of~$B$ is zero, we have again 
$UB=B\hat{U}$ and we can argue as before.
\\
4) Finally we have to consider the case where the matrix $U$ adds a non constant multiple
of one row of~$G$ to another. 
Without restriction we may assume that $z^l$ times the $j$th row is added to the $i$th row.
Since~$G'$ is supposed to be minimal again, we have $l\leq \nu_i-\nu_j$.
Hence $U=I_k+E$ where $E\in\F^{k\times k}$ has the entry $z^l$ at position~$(i,j)$ 
and~$0$ elsewhere.
Let again first $\nu_j>0$.
Consider~$\hat{U}$ as in\eqnref{e-Uhat} but where~$M$ now is of the form
\[
   M=\begin{pmatrix}0_{l\times\nu_j}\\I_{\nu_j}\\0\end{pmatrix}\in\F^{\nu_i\times\nu_j}.
\]
Then we obtain
\[
  A'=A,\ B'=B,\ C'=\hat{U}C+\hat{E}D,\,D'=D
\]
where $\hat{E}\in\F^{\gamma\times k}$ has a~$1$ at position~$(r,j)$ with 
$r=\sum_{\tau=1}^{i-1}\nu_{\tau}+l$ and~$0$ elsewhere.
Furthermore, $A_{i}M=MA_j+N$ where $N\in\F^{\nu_i\times\nu_j}$ has a~$1$ at position~$(l,1)$ 
and~$0$ elsewhere.
Thus $A\hat{U}=\hat{U}A+\hat{N}$ where $\hat{N}\in\F^{\gamma\times\gamma}$ satisfies
$\hat{N}_{r,t}=1$ with~$r$ as above and $t=\sum_{\tau=1}^{j-1}\nu_{\tau}+1$ and all other 
entries are zero.
Moreover, one has $\hat{E}B=\hat{N}$, since $\nu_j>0$, as well as $B\hat{U}=B$ since $l>0$.
Suppose now that $X\edge{u}{v}X'$ is an edge in the state diagram of~$G$. Then 
$X'=XA+uB$ and $v=XC+uD$. One computes
\[
  X'\hat{U}^{-1}=(X\hat{U}^{-1})A'+(u-X\hat{U}^{-1}\hat{E})B'\text{ and }
  v=(X\hat{U}^{-1})C'+(u-X\hat{U}^{-1}\hat{E})D'.
\]
Thus, putting $\tilde{u}=u-X\hat{U}^{-1}\hat{E}$, we obtain that
$X\hat{U}^{-1}\edge{\tilde{u}}{v}X'\hat{U}^{-1}$ is an edge of the state diagram of~$G'$.
Again $\hat{U}^{-1}$ simply permutes the states and we obtain 
$\Pi\Lambda_G\Pi^{-1}=\Lambda_{G'}$ for a suitable permutation~$\Pi$.
In the case $\nu_j=0$ we have $A'=A,\,B'=B,\,C'=C+\hat{E}D$, and $D'=D$ where~$\hat{E}$ is 
as before.
Since the $j$th row of~$B$ is zero, one has $\hat{E}B=0$ and thus the equations
$X'=XA+uB,\,v=XC+uD$ are equivalent to the equations
$X'=XA'+(u-X\hat{E})B',\,v=XC'+(u-X\hat{E})D'$ and we can argue as above.
This completes the proof.
\end{proof}
Notice that the proof also shows how the controller canonical form changes under
unimodular transformations of the minimal generator matrix. 
However, we will not need that result explicitly.

Next we will briefly turn to monomially equivalent codes.
Defining monomial equivalence just like for block codes it is straightforward to show that it preserves
the generalized adjacency matrices, see Theorem~\ref{T-strEquiv} below. 
In the next section we will see that for certain classes of codes even the converse of that theorem is true.
\begin{defi}\label{D-strEquiv}
Two matrices $G,\,G'\in\F[z]^{k\times n}$ are called {\em monomially equivalent \/}
if $G'=GPR$ for some permutation matrix $P\in Gl_n(\F)$ and a non-singular diagonal 
matrix $R\in Gl_n(\F)$. 
Thus, $G$ and~$G'$ are monomially equivalent if and only if they differ
by a permutation and a rescaling of the columns.
We call two codes $\cC,\,\cC'\subseteq\Flaurent^n$ 
{\em monomially equivalent\/} and write $\cC\sim\cC'$ if $\cC=\im G$ and $\cC'=\im G'$ 
for some monomially equivalent generator matrices.
Furthermore, we write 
\[
  [\cC]:=\{\cC'\subseteq\Flaurent^n\mid \cC'\sim\cC\}
\]
for the monomial equivalence class of the code $\cC\subseteq\Flaurent^n$.
\end{defi}
It is clear that monomially equivalent codes have the same distance.
As we show next they even have the same generalized adjacency matrix. 

\begin{theo}\label{T-strEquiv}
Let $\cC,\,\cC'\subseteq\Flaurent^n$ be two convolutional codes. Then
\[
  \cC\sim\cC'\Longrightarrow \bar{\Lambda}(\cC)=\bar{\Lambda}(\cC').
\]
\end{theo}
\begin{proof}
Let $G,\,G'\in\F[z]^{k\times n}$ be generator matrices of $\cC$ and~$\cC'$, 
respectively, such that $G'=GPR$ for some permutation matrix~$P$ and a nonsingular 
constant diagonal matrix~$R$.
Then the controller canonical forms satisfy $A'=A,\,B'=B,\,C'=CPR,D'=DPR$. 
Thus, if $X\edge{u}{v}X'$ is an edge in the state diagram associated with~$G$ then 
$X\edge{u}{vPR}X'$ is an edge in the state diagram 
of~$G'$. Since $\wt(v)=\wt(vPR)$ we obtain $\Lambda_G=\Lambda_{G'}$ if we fix for both 
codes the same ordering on the state space. This implies the desired result.
\end{proof}

As a consequence we have a well-defined mapping
\begin{equation}\label{e-Lambdahat}
\begin{split}
  \hat{\Lambda}:\,\big\{[\cC]\,\big|\,\cC\subseteq\Flaurent^n
     \text{ code with overall constraint length }\gamma\big\}
     &\longrightarrow\Q[W]^{q^{\gamma}\times q^{\gamma}}/_{\textstyle\cG}\\
   [\cC]\hspace*{3cm}&\longmapsto \qquad \bar{\Lambda}(\cC)
\end{split}
\end{equation}
In the next section we will discuss some properties of the mappings $\bar{\Lambda}$ and
$\hat{\Lambda}$.

\section{The adjacency matrix as a complete invariant for one-dimen\-sion\-al binary codes}
\label{S-compl.inv}
\setcounter{equation}{0}
In this section we will derive some results about the generalized adjacency matrix. 
The guiding question is as to what properties do codes share if they have the same 
generalized adjacency matrix. 
We will first show that such codes always have the same Forney indices. 
Secondly, we will show that binary one-dimensional codes with the same generalized adjacency matrix 
are monomially equivalent.  
We conjecture that this is also true for one-dimensional codes over
arbitrary fields.
However, it is certainly not true for general higher-dimensional codes as we know from (binary) 
block code theory, see \cite[Exa.~1.6.1]{HP03}.
As a consequence, we obtain that if two binary one-dimensional codes share the same generalized adjacency 
matrix, then so do their duals.
This indicates the existence of a MacWilliams duality theorem for the adjacency matrices of convolutional 
codes, and, indeed, such a theorem has been proven for codes with overall constraint length one in the
paper~\cite{Ab92}, see\eqnref{e-MacWtrafo} at the end of this section.
The general case has to remain open for future research.

We begin with showing that two codes sharing the same generalized adjacency matrix have 
the same Forney indices.
Observe that two such codes certainly have the same overall constraint length since
that determines the size of the adjacency matrix. 
\begin{theo}\label{T-sameindices}
Let $\cC,\,\cC'\subseteq\Flaurent^n$ be two convolutional codes with the same overall 
constraint length and such that
$\bar{\Lambda}(\cC)=\bar{\Lambda}(\cC')$. 
Then $\cC$ and $\cC'$ have the same dimension and, up to ordering, the same Forney 
indices.
\end{theo}
For the proof we will need the following lemma.

\begin{lemma}\label{L-restrsim}
Let the data be as in General Assumption~\ref{GA}.
Let $\Lambda_G$ be the adjacency matrix of~$G$ with respect to some fixed ordering
on the state space~$\F^{\gamma}$ such that $X_0=0$.
Put $\Gamma:=\Lambda_G+E_{0,0}$ where $E_{0,0}\in\Q^{s\times s}$ is the matrix 
with~$1$ at position $(0,0)$ and~$0$ elsewhere.
Define
\[
   \rho_r=\rank\begin{pmatrix}B\\BA\\ \vdots\\ BA^r\end{pmatrix}
\]
for $r\in\N_0$.
Then for $r\geq1$ the number of nonzero entries in the first row of $\Gamma^r$ is given 
by $q^{\rho_{r-1}}$.
\end{lemma}
The matrix $\Gamma$ is the adjacency matrix of the state diagram where also the edge $0\edge{0}{0}0$ 
is included. 
We call this the extended state diagram.
\\[.6ex]
\begin{proof}
Just like in the proof of Proposition~\ref{P-Lambda1} one shows that 
$\Gamma^r_{i,j}$ is the
weight enumerator of all paths from $X_i$ to~$X_j$ of length exactly~$r$ in the 
extended state diagram. 
Thus our assertion is equivalent to saying that there are exactly $q^{\rho_{r-1}}$ 
states that can be reached from $X_0=0$ by a path of length exactly~$r$ in the 
extended state diagram. 
This is obviously true for $r=1$ since the existence of an edge $0\edge{u}{v}X_j$
is equivalent to the existence of~$u$ such that $X_j=uB$ and there are
$q^{\text{rk}B}=q^{\rho_0}$ different $X_j$ possible. 
In general, the existence of a path
\[
  0\edge{u_1}{v_1}X_{j_1}\edge{u_2}{v_2}\ldots\edge{u_r}{v_r}X_{j_r}
\]
is equivalent to the existence of $u_1,\ldots,u_r$ such that
\[
  X_{j_r}=(u_r,u_{r-1},\ldots,u_1)
  \begin{pmatrix}B\\ BA\\ \vdots\\ BA^{r-1}\end{pmatrix}
\]
showing that this allows for $q^{\rho_{r-1}}$ different states $X_{j_r}$.
\end{proof}

Now it is not hard to prove the theorem above. 

\noindent{\sc Proof of Theorem~\ref{T-sameindices}:}
Let $G\in\F[z]^{k\times n}$ and $G'\in\F[z]^{k'\times n}$ be minimal generator matrices of~$\cC$ 
and~$\cC'$ and $(A,B,C,D)$ and $(A',B',C',D')$ the controller canonical forms, respectively. 
Put $\Gamma=\Lambda_G+E_{0,0}$ and $\Gamma'=\Lambda_{G'}+E_{0,0}$ where $E_{0,0}$ is 
as in the previous lemma.
By assumption $\Pi\Lambda_G\Pi^{-1}=\Lambda_{G'}$ for some $\Pi\in\cG$.
Then also $\Pi\Gamma\Pi^{-1}=\Gamma'$, since $\Pi_{0,0}=1$. 
Thus the first rows of~$\Gamma$ and~$\Gamma'$ coincide up to ordering.
From this we obtain $k=k'$.
Indeed, the first row of $\Gamma$ contains the weight of all edges emerging from the zero state 
in the extended state diagram associated with~$G$. 
Since there are $q^k$ such edges we have $\sum_{j=0}^{s-1}\Gamma_{0,j}=\sum_{\alpha=0}^na_{\alpha}W^{\alpha}$ 
where $\sum_{\alpha=0}^na_{\alpha}=q^k$.
On the other hand 
$\sum_{j=0}^{s-1}\Gamma_{0,j}=\sum_{j=0}^{s-1}\Gamma_{0,j}'$
and using the same argument this yields $k=k'$.
As for the Forney indices we proceed as follows.
We have $\Pi\Gamma^r\Pi^{-1}=(\Gamma')^r$ for all $r\in\N$.
Thus, due to the form of $\Pi$, the matrices $\Gamma^r$ and $\Gamma'^r$ 
have the same number of nonzero entries in the first row.
Therefore, Lemma~\ref{L-restrsim} implies $\rho_r=\rho'_r$ for all $r\in\N_0$ where 
\[
  \rho_r=\rank\begin{pmatrix}B\\BA\\ \vdots\\BA^r\end{pmatrix},\quad
  \rho'_r=\rank\begin{pmatrix}B'\\B'A'\\ \vdots\\B'A'^r\end{pmatrix}.
\]
Now let $\gamma_1,\ldots,\gamma_k$ and $\gamma'_1,\ldots,\gamma'_k$ be the Forney indices of~$G$ and~$G'$, 
respectively. 
By definition of~$A$ and~$B$ we have
\[
  \rho_0=\rank B=\#\{i\mid \gamma_i>0\} \text{ and }
  \rank BA^r =\#\{i\mid \gamma_i>r\}\text{ for all }r\in\N.
\]
Moreover, due to the specific form of the matrices,
\[
   \rho_r=\rank\begin{pmatrix}B\\BA\\ \vdots\\BA^{r-1}\end{pmatrix}+\rank BA^r.
\]
Therefore, 
\[
    \rho_r-\rho_{r-1}=\#\{i\mid \gamma_i>r\}\text{ for }r\in\N.
\]
Analogous identities hold true for the other code.
Using now $\rho_r=\rho_r'$ for all $r\in\N_0$ it follows
\[
  \#\{i\mid \gamma_i>r\}=\#\{i\mid \gamma'_i>r\}\text{ for all }r\in\N_0.
\]
This shows that the Forney indices coincide up to ordering.
\hfill$\Box$

Now we come to the main result of this section.
The proof will make use of the Equivalence Theorem of MacWilliams about weight 
preserving transformations for block codes. 
Moreover, a technical lemma for bijections on $\F_2^{\gamma}$ will be proven in the 
appendix. 
Even though the lemma does not hold true for arbitrary fields we strongly believe that the 
following theorem is also valid for one-dimensional codes over bigger fields.
\begin{theo}\label{T-strEquiv2}
Let $\cC,\,\cC'\subseteq\F_2(\!(z)\!)^n$ be two binary one-dimensional codes such that
$\bar{\Lambda}(\cC)=\bar{\Lambda}(\cC')$. 
Then $\cC$ and $\cC'$ are monomially equivalent.
\end{theo}
Notice that the result implies that the mapping~$\hat{\Lambda}$ in\eqnref{e-Lambdahat}
restricted to binary one-dimensional codes is bijective.
One should also observe that the theorem above is not true for higher-dimensional binary codes 
since it even fails for binary block codes, see \cite[Exa.~1.6.1]{HP03}.
\\[1ex]
\begin{proof}
Let $\F=\F_2$ and~$G,\,G'\in\F[z]^{1\times n}$ be minimal generator matrices of~$\cC$ 
and~$\cC'$, respectively. 
By assumption $\cC$ and~$\cC'$ have the same overall constraint length, say~$\gamma$.
Without loss of generality we may assume $\gamma\geq1$.
Fix an ordering on the state space~$\F^{\gamma}$ such that $X_0=0$ and let $\Lambda_G,\,\Lambda_{G'}$ 
be the adjacency matrices associated with~$G$ and~$G'$.
By assumption $\Pi\Lambda_{G'}\Pi^{-1}=\Lambda_G$ for
some $\Pi\in\cG$.
If $\gamma=1$ then $\cG=\{I_2\}$, thus $\Pi=I_2$.
Next we will show that also in the case $\gamma\geq2$ we obtain $\Pi=I_{2^{\gamma}}$.
In order to do so, let
$S_{2^{\gamma}}$ be the symmetric group on the set $\{0,1,\ldots,2^{\gamma}-1\}$ and define
$\pi\in S_{2^{\gamma}}$ to be the permutation such that
\[
  \Pi=\begin{pmatrix}e_{\pi(0)}\\e_{\pi(1)}\\ \vdots\\ e_{\pi(2^\gamma-1)}\end{pmatrix},
  \quad \Pi^{-1}=\big(e_{\pi(0)}^{\sf T},\,e_{\pi(1)}^{\sf T},\,\ldots,\,
                      e_{\pi(2^{\gamma}-1)}^{\sf T}\big),
\]
where $e_0,\ldots,e_{2^{\gamma}-1}$ are the standard basis vectors in $\Q^{2^{\gamma}}$.
Then $\pi(0)=0$.
The controller canonical forms of~$G$ and~$G'$ are given by $(A,B,C,D)$ and 
$(A,B,C',D')$ where
\[
  A=\begin{pmatrix}\ &1&\ &\  \\ \ & &1&  \\ & & &\ddots & \\ & & & &1\\ & & & &
     \end{pmatrix}\in\F^{\gamma\times\gamma},\quad
  B=\big(1,0,\ldots,0\big)\in\F^{\gamma}
\]
and $C,D,C',D'$ are defined as in Definition~\ref{D-ABCD}.
We have $(\Lambda_G)_{i,j}=W^{\alpha}$ if and only if there exists $u\in\F$ such that
$X_i\edge{u}{v}X_j$ is an edge in the state diagram of~$G$ and $\wt(v)=\alpha$. 
In the rest of the proof we will use for $X=(x_1,\ldots,x_{\gamma})\in\F^{\gamma}$ and 
$1\leq a\leq b\leq\gamma$ the notation $X_{[a,b]}:=(x_a,\ldots,x_b)$.
Then the existence of the edge $X_i\edge{u}{v}X_j$ is equivalent to $X_j=(u,(X_i)_{[1,\gamma-1]})$ 
and, in particular,~$u$ is uniquely determined by~$X_j$.
On the other hand all this is equivalent to 
$(\Lambda_{G'})_{\pi(i),\pi(j)}=(\Pi\Lambda_{G'}\Pi^{-1})_{i,j}=(\Lambda_G)_{i,j}=W^{\alpha}$,
hence to the existence of an edge $X_{\pi(i)}\edge{u'}{v'}X_{\pi(j)}$ such that 
$\wt(v')=\alpha$ in the state diagram of~$G'$.
Likewise we have $X_{\pi(j)}=(u',(X_{\pi(i)})_{[1,\gamma-1]})$.
Denote by $\hat{\pi}:\F^{\gamma}\longrightarrow\F^{\gamma}$ the permutation on $\F^{\gamma}$
such that $\hat{\pi}(X_i)=X_{\pi(i)}$ for all $i=0,\ldots,2^{\gamma}-1$.
Then $\hat{\pi}(0)=0$ and the above gives us the following property for the permutation 
$\hat{\pi}$:
\[
  Y_{[2,\gamma]}=X_{[1,\gamma-1]}\Longrightarrow
  \hat{\pi}(Y)_{[2,\gamma]}=\hat{\pi}(X)_{[1,\gamma-1]}
  \text{ for all }X,\,Y\in\F^{\gamma}.
\]
In Lemma~\ref{L-permut} in the appendix we show that this implies $\hat{\pi}=\text{ id}$.
Thus $\Pi=I_{2^{\gamma}}$.
As a consequence we have $\Lambda_{G'}=\Lambda_G$ for all $\gamma\geq1$.
From the equivalence
\[
   (\Lambda_G)_{i,j}=W^{\alpha}\Longleftrightarrow \wt(X_iC+X_{j,1}D)=\alpha
\]
and the corresponding equivalence for $(\Lambda_{G'})_{i,j}=W^{\alpha}$ we finally arrive at
\[
  \wt\Big((X_i,u)\begin{pmatrix}C\\D\end{pmatrix}\Big)
  =\wt\Big((X_i,u)\begin{pmatrix}C'\\D'\end{pmatrix}\Big)
  \text{ for all }X_i\in\F^{\gamma},\,u\in\F.
\]
But then Lemma~\ref{L-MacW} below yields 
\[
  \begin{pmatrix}C'\\D'\end{pmatrix}=\begin{pmatrix}C\\D\end{pmatrix}P
\]
for some permutation matrix $P\in Gl_n(\F)$.
Hence $G'=GP$ meaning that the two matrices are monomially equivalent. 
\end{proof}

It remains to prove 
\begin{lemma}\label{L-MacW}
Let $\F$ be any finite field and let $M,\,M'\in\F^{k\times n}$ be such that 
\begin{equation}\label{e-sameweight}
   \wt(uM)=\wt(uM') \text{ for all }u\in\F^k.
\end{equation}
Then $M$ and $M'$ are monomially equivalent.
\end{lemma}
\begin{proof}\footnote{One can prove this result straightforwardly for the 
field~$\F_2$. 
However, I wish to thank Gert Schneider for pointing out the connection to
MacWilliams' Equivalence Theorem to me.}
Let us first assume that $M$ has rank~$k$. Then the assumption\eqnref{e-sameweight}
implies that $M'$ has rank~$k$, too.
Defining the block codes $\cB=\im M$ and $\cB'=\im M'$ 
we obtain a well-defined weight-preserving bijective linear transformation
\[
   \cB\longrightarrow \cB',\quad uM\longmapsto uM'.
\]
By virtue of the Equivalence Theorem of MacWilliams, see for instance 
\cite[Thm.~7.9.4]{HP03} the two block codes are monomially equivalent. 
Thus there exist a permutation matrix~$P$
and a non-singular diagonal matrix~$R$ in $Gl_n(\F)$ such that $M'=MPR$.
\\
Let now $\rank M=r<k$ and assume without loss of generality that 
$M=\Smalltwomat{M_1}{0}$ where $M_1\in\F^{r\times n}$ has full row rank.
Then $(0,u_2)M=0$ for all $u_2\in\F^{k-r}$ and\eqnref{e-sameweight} implies that
$M'=\Smalltwomat{M_1'}{0}$ for some $M_1'\in\F^{r\times n}$ with full row rank.
Now we have $\wt(u_1M_1)=\wt(u_1M_1')$ for all $u_1\in\F^r$ and by the first part of this
proof~$M_1$ and~$M_1'$ are monomially equivalent. But then the same is true for~$M$ 
and~$M'$.
\end{proof}

We close the section with briefly discussing the question whether there might exist a
MacWilliams duality theorem for convolutional codes.
For block codes this famous theorem states that the weight distribution of the dual code is fully 
determined by the weight distribution of the original code and a transformation formula 
is given, see, e.~g.,~\cite[Thm.~3.5.3]{vLi99}.
For convolutional codes one might think of two possible generalizations of this result, either to the
weight distribution~$\Omega$ or to the adjacency matrix~$\Lambda$. 
As we will describe next, both cases have already been touched upon in the literature.
In~\cite{SM77} it has been shown that there does not exist a MacWilliams duality theorem for 
the weight distribution~$\Omega$ of convolutional codes.
Precisely, the following example has been presented.
Consider $G_1=[1,\,z,\,1+z],\,G_2=[z,\,z,\,1+z]\in\F_2^{1\times3}$.
Then one shows that the weight distributions of the two codes $\cC_1:=\im G_1$ and 
$\cC_2=\im G_2$ coincide. 
Indeed, they are both given by $\Omega=\frac{L^2W^4}{1-LW^2}$.
The dual codes are given by 
\[
   \cC_1^{\perp}=\im\begin{pmatrix}1&1&1\\z&1&0\end{pmatrix},\
   \cC_2^{\perp}=\im\begin{pmatrix}1&1&0\\1+z&0&z\end{pmatrix},
\]
and it turns out that they have different weight distributions
\[
  \Omega_{\cC_1^{\perp}}=\frac{L^2W^2+LW^3+2L^2W^3-L^2W^5}{1-LW-LW^2},\
  \Omega_{\cC_2^{\perp}}=\frac{LW^2+3L^2W^3-L^2W^5}{1-LW-LW^3}.
\]
As a consequence there cannot exist a MacWilliams transformation mapping the weight 
distribution of a given code onto the weight distribution of the dual without 
using any further information. 
The example even shows more. Since multiplication by~$z$ is weight-preserving, the mapping 
$uG\longmapsto uG'$ yields an isometry between the codes~$\cC$ and~$\cC'$. 
But obviously, the codes are not monomially equivalent, showing that there is no MacWilliams
Equivalence Theorem for convolutional codes in this form (one would have to allow at least rescaling 
by powers of~$z$ in monomial equivalence).
Let us now discuss the adjacency matrices of these codes. 
Since the two codes are not monomially equivalent we know from
Theorem~\ref{T-strEquiv2} that the generalized adjacency matrices of the two codes
are not identical. Indeed, one computes
\[
  \Lambda_1=\begin{pmatrix}0&W^2\\W^2&W^2\end{pmatrix},\
  \Lambda_2=\begin{pmatrix}0&W\\W^3&W^2\end{pmatrix}.
\]
Of course, the generalized adjacency matrices of the dual codes are different since the 
weight distributions are. 
They are given by
\[
  \Lambda_1^{\perp}=\begin{pmatrix}W^3&W+W^2\\W+W^2&W+W^2\end{pmatrix},\
  \Lambda_2^{\perp}=\begin{pmatrix} W^2&2W\\2W^2&W+W^3\end{pmatrix}.
\]
At this point the question arises whether there exists a MacWilliams duality theorem for the 
adjacency matrices of convolutional codes. 
Indeed, in the paper~\cite{Ab92} such a transformation has been established for codes with overall
constraint length one. 
It is derived in totally different notation, but for codes with parameters $(n,k,1)_2$ it amounts 
after some rewriting to the formula
\begin{equation}\label{e-MacWtrafo}
  \Gamma^{\perp}=2^{-k-1}(1+W)^nM\T|_{\frac{1-W}{1+W}}\text{ where }
  M=\begin{pmatrix}1&1\\1&-1\end{pmatrix}\Gamma\begin{pmatrix}1&1\\1&-1\end{pmatrix},
\end{equation}
and where $M|_{a}$ denotes substitution of~$a$ for~$W$ in every entry of the matrix~$M$ and 
$\Gamma=\Lambda+E_{0,0}$ is the adjacency matrix of the extended state diagram 
(see also Lemma~\ref{L-restrsim}).
The formula can straightforwardly be verified for the two codes and their duals given above.

We strongly believe that such a transformation also exists for codes with bigger overall constraint length.
At least in the one-dimensional binary case with arbitrary overall constraint length we can establish 
the following support for this conjecture.
\begin{cor}\label{C-dual}
Let $\cC,\,\cC'\subseteq\F_2(\!(z)\!)^n$ be two binary one-dimensional codes.
Then 
\[
   \bar{\Lambda}(\cC)=\bar{\Lambda}(\cC')\Longrightarrow
   \bar{\Lambda}(\cC^{\perp})=\bar{\Lambda}(\cC'^{\perp}).
\]
\end{cor}
\begin{proof}
By virtue of Theorem~\ref{T-strEquiv2} the assumption implies that $\cC$ and~$\cC'$ 
are monomially equivalent.
It is trivial to see that then also $\cC^{\perp}$ and $\cC'^{\perp}$ are monomially equivalent 
and thus Theorem~\ref{T-strEquiv} yields 
$\bar{\Lambda}(\cC^{\perp})=\bar{\Lambda}(\cC'^{\perp})$.
\end{proof}
Unfortunately, the corollary does not reveal a formula transforming $\bar{\Lambda}(\cC)$ into
$\bar{\Lambda}(\cC^{\perp})$.

We close the section with the following 
\\[1ex]
{\bf Conjecture: }
Let $\cC,\,\cC'\subseteq\F(\!(z)\!)^n$ be two codes.
Then 
\[
   \bar{\Lambda}(\cC)=\bar{\Lambda}(\cC')\Longrightarrow
   \bar{\Lambda}(\cC^{\perp})=\bar{\Lambda}(\cC'^{\perp}).
\]

\section{Open Problems}\label{S-OP}
With this paper we want to initiate an investigation of the weight distribution and weight preserving 
maps for convolutional codes.
The central object of our approach is the adjacency matrix of the associated state diagram.
In Theorem~\ref{T-strEquiv2} we showed that for one-dimensional binary codes this matrix
uniquely determines the code up to monomial equivalence. 
One immediately wonders whether this is true for one-dimensional codes over arbitrary fields as well. 
From block code theory it is known that such a result, however, cannot be expected for 
{\em higher-dimensional\/} codes.
It would be helpful to see some examples with positive overall constraint length.
Moreover, it needs to be investigated if isometries between convolutional codes can be 
described explicitly. 
Finally, of course there remains the conjecture at the end of the last section that the adjacency 
matrix of a given code determines that of the dual code. 
While for codes with overall constraint length one a transformation has been derived in~\cite{Ab92},
the general case has to remain open for future research.

\appendix
\section*{Appendix}
\setcounter{section}{1}
\setcounter{theo}{0}
\renewcommand{\theequation}{A.\arabic{equation}}
\setcounter{equation}{0}

In the following lemma we use again the notation 
$X_{[a,b]}:=(X_a,X_{a+1},\ldots,X_{b})$ for 
$X=(X_1,\ldots,X_{\gamma})\in\F^{\gamma}$ and all $1\leq a\leq b\leq\gamma$.
For $X_{[a,a]}$ we write, of course, simply $X_a$.
\begin{lemma}\label{L-permut}
Let $\F=\F_2$ and $\gamma\geq2$. 
Furthermore, let $\hat{\pi}:\F^{\gamma}\longrightarrow\F^{\gamma}$ be a bijective map 
with $\hat{\pi}(0)=0$ and satisfying
\begin{equation}\label{e-shiftPi}
   \hat{\pi}(u,X_{[1,\gamma-1]})_{[2,\gamma]}=\hat{\pi}(X)_{[1,\gamma-1]}
   \text{ for all }X\in\F^{\gamma}\text{ and all }u\in\F.
\end{equation}
Then~$\hat{\pi}$ is the identity map.
\end{lemma}
\begin{proof}
Denote by $e_1,\ldots,e_{\gamma}$ the standard basis vectors on $\F^{\gamma}$.
\\
1) Using $X=0$ and $u=1$ we obtain 
$\hat{\pi}(e_1)_{[2,\gamma]}=\hat{\pi}(0)_{[1,\gamma-1]}=(0,\ldots,0)$, thus 
$\hat{\pi}(e_1)=(a,0,\ldots,0)$.
Bijectivity of $\hat{\pi}$ implies $a=1$, thus $\hat{\pi}(e_1)=e_1$.
\\
2) Using $X=e_{\gamma}$ and $u=0$ we obtain
\[
  \hat{\pi}(u,X_{[1,\gamma-1]})_{[2,\gamma]}=\hat{\pi}(0)_{[2,\gamma]}=0=
  \hat{\pi}(e_{\gamma})_{[1,\gamma-1]},
\]
thus $\hat{\pi}(e_{\gamma})=(0,\ldots,0,a)$ and again bijectivity of~$\hat{\pi}$ implies 
$\hat{\pi}(e_{\gamma})=e_{\gamma}$.
\\
3) Now we proceed by induction. Assume that there is some $r\geq 1$ such that
$\hat{\pi}(X)=X$ for all $X\in\F^{\gamma}$ satisfying $\wt(X)\leq r$ and $X_1=1$.
By~1) this is true for $r=1$.
Then we have to show
\begin{romanlist}
\item $\hat{\pi}(\tilde{X})=\tilde{X}$ for all~$\tilde{X}$ such that $\wt(\tilde{X})\leq r$,
\item $\hat{\pi}(\tilde{X})=\tilde{X}$ for all~$\tilde{X}$ such that 
      $\wt(\tilde{X})\leq r+1$ and $\tilde{X}_1=1$.
\end{romanlist}
\underline{Ad~(i):}
Pick $X\in\F^{\gamma}$ such that $\wt(X)\leq r$ and $X_1=1$. 
Put $X^{(1)}=(0,X_{[1,\gamma-1]})$.
Then $\hat{\pi}(X^{(1)})_{[2,\gamma]}=\hat{\pi}(X)_{[1,\gamma-1]}=X_{[1,\gamma-1]}$, thus
$\hat{\pi}(X^{(1)})=(a_1,X_{[1,\gamma-1]})$.
Put now $X^{(i)}=(0,\ldots,0,X_{[1,\gamma-i]})\in\F^{\gamma}$. 
We proceed by induction on~$i$. 
Thus by hypothesis we may assume 
\begin{equation}\label{e-indhypo}
  \hat{\pi}(X^{(i)})=(a_i,\ldots,a_1,X_{[1,\gamma-i]}).
\end{equation}
Then $X^{(i+1)}=(0,X^{(i)}_{[1,\gamma-1]})$ and thus
\[
  \hat{\pi}(X^{(i+1)})_{[2,\gamma]}=\hat{\pi}(X^{(i)})_{[1,\gamma-1]}
  =(a_i,\ldots,a_1,X_{[1,\gamma-i-1]}).
\]
Therefore $\hat{\pi}(X^{(i+1)})=(a_{i+1},\ldots,a_1,X_{[1,\gamma-i-1]})$.
Hence\eqnref{e-indhypo} holds true for all $i=1,\ldots,\gamma-1$. 
Now $X^{(\gamma-1)}=e_{\gamma}$.
Hence by~2) of this proof
$e_{\gamma}=\hat{\pi}(X^{(\gamma-1)})=(a_{\gamma-1},\ldots,a_1,X_1)$.
This implies $a_1=\ldots=a_{\gamma-1}=0$ and hence $\hat{\pi}(X^{(i)})=X^{(i)}$ for all
$i=1,\ldots,\gamma-1$.
Since each $\tilde{X}\in\F^{\gamma}$ such that $\wt(\tilde{X})\leq r$ is of the form 
$X^{(i)}$ for a suitable $X$ satisfying $\wt(X)\leq r$ and $X_1=1$, this proves~(i).
\\
\underline{Ad~(ii):}
Let $\tilde{X}\in\F^{\gamma}$ such that $\tilde{X}_1=1$ and $\wt(\tilde{X})\leq r+1$. 
Then $\tilde{X}=(1,X_{[1,\gamma-1]})$ for some $X\in\F^{\gamma}$ such that 
$\wt(X)=\wt(X_{[1,\gamma-1]})\leq r$.
By part~(i) we know that $\hat{\pi}(X)=X$ as well as 
\begin{equation}\label{e-Pi1}
   \hat{\pi}(0,X_{[1,\gamma-1]})=(0,X_{[1,\gamma-1]}).
\end{equation}
Now\eqnref{e-shiftPi} yields 
\[
  \hat{\pi}(\tilde{X})_{[2,\gamma]}=\hat{\pi}(X)_{[1,\gamma-1]}=X_{[1,\gamma-1]}.
\]
Hence $\hat{\pi}(\tilde{X})=(a,X_{[1,\gamma-1]})$ and bijectivity of~$\hat{\pi}$ along 
with\eqnref{e-Pi1} yields $a=1$. Thus $\hat{\pi}(\tilde{X})=\tilde{X}$.
\end{proof}

\bibliographystyle{abbrv}
\bibliography{literatureAK,literatureLZ}
\end{document}